\documentclass[onecolumn]{elsart}
\usepackage{amssymb}
\usepackage{amsfonts}
\usepackage{amsmath}
\usepackage{graphicx}
\usepackage{subfigure}
\usepackage{epsfig}
\usepackage{hyperref}
\usepackage{appendix}
\usepackage{cite}

\input{tcilatex}

\begin{document}

\begin{frontmatter}%

\title{Structure of Exact Renormalization Group Equations for field theory}

\author{C.\ Bervillier}\ead{claude.bervillier@lmpt.univ-tours.fr}%

\address{Laboratoire de Math\'{e}matiques et Physique Th\'{e}orique,\\ UMR 7350 (CNRS),\\
F\'ed\'eration Denis Poisson,\\
Universit\'{e} Fran\c{c}ois Rabelais,\\
Parc de Grandmont, 37200 Tours, France}%

\begin{abstract}
It is shown that exact renormalization group (RG) equations (including
rescaling and field-re\-nor\-ma\-li\-za\-tion) for respectively the
scale-de\-pen\-dent \textsl{full} action $S\left[ \tilde{\phi},t\right] $ and
the scale-de\-pen\-dent \textsl{full} effective action $\Gamma \left[ \tilde{\Phi%
},t\right] $ --in which $t$ is the \textquotedblleft
RG-time\textquotedblright\ defined as the logarithm of a running momentum
scale-- may be linked together by a Legendre transformation as simple as $%
\Gamma \left[ \tilde{\Phi},t\right] -S\left[ \tilde{\phi},t\right] \ +\tilde{%
\phi}\cdot \tilde{\Phi}=0$, with $\tilde{\Phi}\left( x\right) =\delta S\left[
\tilde{\phi}\right] /\delta \tilde{\phi}\left( x\right) $ (resp. $\tilde{\phi%
}\left( x\right) =-\delta \Gamma \left[ \tilde{\Phi}\right] /\delta \tilde{%
\Phi}\left( x\right) $), where $\tilde{\phi}$ and $\tilde{\Phi}$ are
dimensionless-renormalized quantities. This result, in which any explicit
reference to a \textquotedblleft cutoff procedure\textquotedblright\ is
absent, makes sense in the framework of field theory. It may be compared to
the dimensional regularization of the perturbative field theory, in which the
running momentum scale is a pure scale of reference and not a momentum cutoff.
It is built from the Wilson
historic first exact RG equation in which the field-renormalization step is
realized via an operator which is redundant and exactly marginal at a fixed
point, the properties of which are conserved by the Legendre transformation
and which modifies the usual removal of the overall UV cutoff $\Lambda _{0}$
by associating it with the removal of an overall IR cutoff $\mu $. Because
the final equations do not refer to any true cutoff (even for the
scale-de\-pen\-dent $\Gamma $), it reinforces the idea that one may get rid of
the achronistic procedure of \textquotedblleft
regularizing\textquotedblright\ the theory\ via an explicit
\textquotedblleft cutoff function\textquotedblright , procedure\ which is
often seen as an inconvenience to treat modern problems in field theory.
\end{abstract}%


\end{frontmatter}

\section{Introduction}

There are two families of exact\footnote{%
This is the historic appellation chosen in order to distinguish these
equations from approximate realizations of the RG steps. Some authors prefer
naming them non-perturbative or functionnal RG equations. However, the term
\textquotedblleft exact\textquotedblright\ has the merit of clearly
designing the subject of interest. Moreover, nobody would have the idea of
changing the historic appellation \textquotedblleft Renormalization
Group\textquotedblright\ on the grounds that the mathematical notion of
group does not play an essential role in the subject.} renormalization group
(RG) equations (ERGE) according to wether one considers a Wilsonian ERGE 
\cite{440} --for the scale-de\-pen\-dent action $S_{\Lambda }\left[ \phi %
\right] $-- or a Wetterichian ERGE --for the scale-de\-pen\-dent effective
action $\Gamma _{k}\left[ M\right] $ (also named \textquotedblleft effective
average action\textquotedblright\ \cite{4374}). The two scales $\Lambda $
and $k$ have the following meanings. When usually referring to an action $S%
\left[ \phi \right] $ and an effective action $\Gamma \left[ M\right] $, $%
\phi \left( x\right) $ represents a scalar field attached to the microscopic
description of a system (like the spin for the Ising model) and $M\left(
x\right) $ a macroscopic quantity, averaged over the whole volume (like the
magnetization). There are, implicitly, two fixed and different scales of
reference (microscopic and macroscopic).\ Instead, with the RG theory, these
scales become continuously variable: $\phi \left( x\right) $ and $M\left(
x\right) $ are both averages over \textsl{partial} volumes of the system,
then $\phi \left( x\right) $ is attached to an ultraviolet (UV) momentum
scale $\Lambda $ (\textquotedblleft short distances\textquotedblright ),
whereas $M\left( x\right) $ is attached to an infrared (IR) momentum scale $%
k $ (\textquotedblleft large distances\textquotedblright ). Despite their
different names, and because they are not fixed, the two scales $k$ and $%
\Lambda $ may well\ be chosen to be equal for a given system. So, in the
following we set $k=\Lambda $.

The RG transformation involves the following three (RG-)steps\footnote{%
In fact, adding the possibility of redefining the field at will provided
that the quasi-locality of the action is not destroyed (see remark \ref%
{Rem:Wil0}), there are four steps allowed (as the three musketeers were in
fact four).}:

\begin{enumerate}
\item[RG-step 1] A reduction of the degrees of freedom (decimation), that
consists of an integration of the high momentum components of the field,
generating a \textquotedblleft scale-de\-pen\-dent action\textquotedblright\
with a reduced \textquotedblleft running\textquotedblright\ scale $\Lambda
\rightarrow \mu <\Lambda $;

\item[RG-step 2] A (classical) rescaling of the momenta back to the
\textquotedblleft initial value of the running scale\textquotedblright , $%
\mu \rightarrow \Lambda $, that is conveniently implemented by imposing that
any dimensioned quantity is rendered dimensionless by means of classical
powers of $\Lambda $ (such as $q=\Lambda \,\tilde{q}$, for a momentum).

\item[RG-step 3] A field-renormalization\footnote{%
It is important to notice that, in the current literature, the
field-renormalization function $Z_{3}$ is most often presented as a function 
$Z_{3}\left( \ell \right) $, with $\ell =\Lambda /\mu _{0}<1$ in which $\mu
_{0}$ is an arbitrary momentum scale introduced for convenience to define
the RG-time $t$ [see eq. (\ref{eq:RGtime})]. In that case the dependence on
the running scale $\Lambda $ within $Z_{3}$ is inverse of the present case
with $\mu /\Lambda <1$. This is because, the current use of $Z_{3}$ in the
construction of an ERGE as a \textsl{global} field-renormalization (i.e.,
over the finite range $\left[ \Lambda ,\mu _{0}\right] $ with $\Lambda <\mu
_{0}$) is introduced as a convenient add-on to the realization of the
RG-steps that are instead themselves effected independently and
infinitesimally via the decreasing $\Lambda \rightarrow \Lambda -d\Lambda $
(see section \ref{Implem}).} $\phi \left( x\right) \rightarrow \sqrt{%
Z_{3}\left( \mu /\Lambda \right) }\,\phi \left( x\right) $ (equivalently $%
M\left( x\right) \rightarrow \sqrt{Z_{3}\left( \mu /\Lambda \right) }%
\,M\left( x\right) $) that removes an indetermination linked to the
reparametrization invariance and potentially introduces the anomalous
dimension of the field $\eta ^{\ast }$ in the vicinity of a fixed point (see
appendix \ref{defZ}).
\end{enumerate}

In principle, an ERGE expresses the evolution of the action (or of the
effective action) under an \textsl{infinitesimal} realization of these three
RG-steps.

Most often one refers to the RG transformation through RG-step 1 exclusively
and one roughly defines an ERGE to be the evolution of $S_{\Lambda }\left[
\phi \right] $\ (resp. $\Gamma _{\Lambda }\left[ M\right] $) under an
infinitesimal change of a momentum cutoff $\Lambda $ artificially introduced
within the actions; the two other RG-steps are usually seen to be secondary
and\ often put into a single step. Actually, we show in this article, that
RG-step 3 may be extremely helpfull in simplifying the relationship between
the two families of ERGE.

One already knows that a Legendre transformation \cite%
{4436,2520,3815,7289,7205,6750} links together RG flow equations for \textit{%
truncations} of both $S_{\Lambda }\left[ \phi \right] $\ and $\Gamma
_{\Lambda }\left[ M\right] $ [noted hereafter respectively $S_{\mathrm{int}%
,\Lambda }\left[ \phi \right] $ and $\Gamma _{\mathrm{int},\Lambda }\left[ M%
\right] $, see (\ref{eq:Sint}) and (\ref{eq:FullGamma0})]. But this Legendre
transformation does not account for RG-steps 2 and 3. When these latter
steps are accounted for, the actions become functions of dimensionless (and
renormalized) quantities noted respectively $S\left[ \tilde{\phi},t\right] $%
\ and $\Gamma \left[ \tilde{M},t\right] $ (similarly $S_{\mathrm{int}}\left[ 
\tilde{\phi},t\right] $ and $\Gamma _{\mathrm{int}}\left[ \tilde{M},t\right] 
$, see section \ref{Nota}) in which $t$ is a dimensionless measure of $%
\Lambda $ called the RG-time defined by (\ref{eq:RGtime}). It is a matter of
fact that the transformation which links $S_{\mathrm{int}}\left[ \tilde{\phi}%
,t\right] $ and $\Gamma _{\mathrm{int}}\left[ \tilde{M},t\right] $ becomes
horribly complicated when the field-renormalization RG-step 3 is accounted
for in an \textquotedblleft ordinary way\textquotedblright , with first
order differential equations to be solved in which the anomalous dimension $%
\eta ^{\ast }$ is very involved \cite{7289,7205,6750}.

In the present paper we show the existence of a Legendre transformation
between the \textit{full} scale-de\-pen\-dent action $S\left[ \tilde{\phi},t%
\right] $\ and the \textit{full} scale-de\-pen\-dent effective action $\Gamma %
\left[ \tilde{\Phi},t\right] $ which is as simple as:%
\begin{eqnarray}
\Gamma \left[ \tilde{\Phi},t\right] -S\left[ \tilde{\phi},t\right] &=&-%
\tilde{\phi}\cdot \tilde{\Phi}\,,  \label{eq:001a} \\
\tilde{\Phi}\left( \tilde{x}\right) &=&\frac{\delta S\left[ \tilde{\phi},t%
\right] }{\delta \tilde{\phi}\left( \tilde{x}\right) }\,,  \label{eq:001b}
\end{eqnarray}%
and which allows to readily deduce one ERGE from the other. This result is
made possible by a particular realization of the
field-re\-nor\-ma\-li\-za\-tion RG-step 3, similar to that originally
utilized by Wilson \cite{440}. It relies on the use of an operator,\ which
is redundant and exactly marginal at fixed points (EMRO), the properties of
which are conserved by the Legendre transformation and which modifies the
consequences of the usual process of sending to infinity the overall
(initial) UV cutoff $\Lambda _{0}>\Lambda $.

An important consequence of (\ref{eq:001a}, \ref{eq:001b}), is that the
fundamental structural properties of the ERGE \cite{4011,2835} usually
expressed exclusively on $S_{\Lambda }\left[ \phi \right] $ may be directly
transposed to $\Gamma _{\Lambda }\left[ M\right] $.

In particular, it is known \cite{4011,2835} that the general structure of an
ERGE for $S_{\Lambda }$ is a consequence of the invariance of the partition
function under an infinitesimal field redefinition (see also \cite{4753} for
the inclusion of the rescaling and field re\-nor\-ma\-li\-za\-tion steps).
Indeed, the following infinitesimal change of field definition:%
\begin{equation*}
\tilde{\phi}_{\tilde{q}}^{\prime }=\tilde{\phi}_{\tilde{q}}-\psi _{\tilde{q}%
}\left( \tilde{\phi}\right) dt\,,
\end{equation*}%
implies the following general expression for a Wilsonian RG flow equation: 
\begin{equation*}
\frac{d}{dt}S\left[ \tilde{\phi},t\right] =\int_{\tilde{q}}\left[ \psi _{%
\tilde{q}}\left( \tilde{\phi}\right) \frac{\delta S\left[ \tilde{\phi},t%
\right] }{\delta \tilde{\phi}_{\tilde{q}}}-\frac{\delta \psi _{\tilde{q}%
}\left( \tilde{\phi}\right) }{\delta \tilde{\phi}_{\tilde{q}}}\right] \,,
\end{equation*}%
in which $\int_{q}\equiv \int \frac{d^{d}q}{\left( 2\pi \right) ^{d}}$ and $%
\tilde{\phi}_{\tilde{q}}$ stands for the Fourier component of the field $%
\tilde{\phi}\left( \tilde{x}\right) $ (for the writing conventions used in
this article, see section \ref{Nota} and \cite{4595})

Now, because it follows from (\ref{eq:001a}, \ref{eq:001b}) that:%
\begin{equation*}
\frac{d}{dt}S\left[ \tilde{\phi},t\right] =\frac{d}{dt}\Gamma \left[ \tilde{%
\Phi},t\right] \,,
\end{equation*}%
then one readily obtains the general expression of the ERGE for $\Gamma %
\left[ \tilde{\Phi},t\right] $: 
\begin{eqnarray*}
\frac{d}{dt}\Gamma \left[ \tilde{\Phi},t\right] &=&\int_{\tilde{q}}\left[
\psi _{\tilde{q}}\left( \tilde{\phi}\right) \,\tilde{\Phi}_{-q}+\frac{1}{%
\Gamma ^{\left( 2\right) }\left( \tilde{q}^{2};\tilde{\Phi}\right) }\frac{%
\delta \psi _{\tilde{q}}\left( \tilde{\phi}\right) }{\delta \tilde{\Phi}_{%
\tilde{q}}}\right] \,, \\
\tilde{\phi}\left( \tilde{x}\right) &=&-\frac{\delta \Gamma \left[ \tilde{%
\Phi},t\right] }{\delta \tilde{\Phi}\left( \tilde{x}\right) }\,,
\end{eqnarray*}%
which may be seen as the consequence of the infinitesimal field redefinition:%
\begin{equation*}
\tilde{\Phi}_{\tilde{q}}^{\prime }=\tilde{\Phi}_{\tilde{q}}+\frac{1}{\Gamma
^{\left( 2\right) }\left( \tilde{q}^{2};\tilde{\Phi}\right) }\psi _{\tilde{q}%
}\left( \tilde{\phi}\right) \,dt\,.
\end{equation*}

Another very interesting property of (\ref{eq:001a}, \ref{eq:001b}) is the
absence of explicit reference to any cutoff function. This enables us to
envisage getting completely rid of the anachronistic necessity of
introducing explicitly within the actions a \textquotedblleft regularization
procedure\textquotedblright\ by a true momentum cutoff which is often seen as an inconvenience in the
framework of field theory.

The organization of the paper is as follows.

In part \ref{Sec:Sum}, we present a review of the various structures of ERGE
(with a smooth cutoff) that are encountered in the literature. We emphasize
both the way RG-step 3 (field-renormalization) is commonly implemented and
the resulting form of the Legendre transformation that links the two
families of ERGE. This part is divided into sections. We begin by a
presentation of our notations (section \ref{Nota}) in which the definition
and properties of the field-renormalization function $Z_{3}$ are recalled.
In section \ref{WilERGE} we present the historic first ERGE proposed by
Wilson \cite{440}. We recall the simple form of the associated EMRO and put
forward its likely role in the original construction of that ERGE together
with the absence of any reference to an explicit UV-cutoff. In section \ref%
{PolERGE}, we present the original Polchinski version \cite{354} written for
a truncation $S_{\mathrm{int},\Lambda }\left[ \phi \right] $ of $S_{\Lambda }%
\left[ \phi \right] $ and without any consideration of RG-steps 2 and 3.
Section (\ref{modif}) shows how these latter two steps have been currently
implemented to give the so-called \textquotedblleft
modified\textquotedblright\ Polchinski ERGE \cite{3491}. Once reexpressed as
a RG flow equation for the full action $S\left[ \tilde{\phi},t\right] $,
this leads to a \textquotedblleft modified\textquotedblright\ Wilsonian ERGE 
\cite{3491} for which the EMRO determined in \cite{6228} takes on a
complicated form (section \ref{EMROWilMod}). In section \ref{WettERGE} we
briefly recall how the ERGE for a truncation $\Gamma _{\mathrm{int},\Lambda }%
\left[ M\right] $ of the effective action $\Gamma _{\Lambda }\left[ M\right] 
$ --most currently known as the Wetterich ERGE (for a review, see \cite{4700}%
)--, may be obtained from the Polchinski ERGE via a Legendre transformation 
\cite{2520,4436}. We then underline the extreme complexity of the most
elaborated Legendre transformation (between $S_{\mathrm{int}}\left[ \tilde{%
\phi},t\right] $ and $\Gamma _{\mathrm{int}}\left[ \tilde{M},t\right] $)
found so far after implementing RG-step 3 ordinarily \cite{7289,7205,6750}
(section \ref{Elab}).

In part \ref{WilsonWay} we propose a structural method for determining the
forms of the various kinds of ERGE listed in part \ref{Sec:Sum} so as to
maintain simple relations between them. The method, inspired by Wilson's
procedure of implementing RG-step 3, is described in section \ref{Prince}.
It is based on the realization of RG-step 3 in the ERGE via an EMRO the
expression of which is extended out of the vicinity of fixed points. Since
the notion of EMRO (seen as a zero-eigenvalue \textquotedblleft
operator\textquotedblright ) exists for any ERGE, we may look at the
evolution of one particular EMRO through the various kinds of ERGE. The
natural starting point is the Wilson ERGE \textquotedblleft \textsl{extended
to an arbitrary cutoff function}\textquotedblright \footnote{%
This is somewhat a misleading expression because there is no actual cutoff
in a Wilsonian ERGE.}\ and its associated still simple EMRO (section \ref%
{Extended}). A new analytical derivation of this equation is given in
section \ref{ana} where it is shown that an effective IR-cutoff $\mu $
appears which is the natural counter part of an overall UV-cutoff $\Lambda
_{0}$. Then it is argued that the recourse to an EMRO necessitates the
renormalization of the field over all the scales and this implies to having
performed the usual limit $\Lambda _{0}\rightarrow \infty $ that
automatically induces the limit $\mu \rightarrow 0$ or reciprocally (section %
\ref{MuVers0}). We then look at the RG-flows for $S_{\mathrm{int}}\left[ 
\tilde{\phi},t\right] $ and for $\Gamma _{\mathrm{int}}\left[ \tilde{M},t%
\right] $ while keeping the natural simple expression of the Legendre
transformation between $S_{\mathrm{int}}\left[ \tilde{\phi},t\right] $ and $%
\Gamma _{\mathrm{int}}\left[ \tilde{M},t\right] $ (section \ref{Struc}).
After having recalled the conditions under which the usual UV limit $\Lambda
_{0}\rightarrow \infty $ is justified (vicinity of a fixed point), we show
that the associated IR limit $\mu \rightarrow 0$ is valid if $\eta ^{\ast }<2
$, a usual condition for a fixed point to be a \textquotedblleft critical
fixed point\textquotedblright\ \ \cite{6699}\ (section \ref{Limite}). The
latter IR limit modifies the relation between the IR and UV cutoff functions
so that a direct and very simple Legendre transformation then relies the
Wilson ERGE \textquotedblleft \textsl{extended to an arbitrary cutoff
function}\textquotedblright\ (for the \textsl{full} action $S\left[ \tilde{%
\phi},t\right] $) to the RG-flow equation for the \textsl{full} effective
action $\Gamma \left[ \tilde{M},t\right] $ that has been structurally
constructed. We conclude in section \ref{Conc}. Because the
field-renormalization plays an essential role in our discussion we also
recall, in appendix \ref{defZ}, how and why $Z_{3}$ is related to the
anomalous dimension of the field $\eta ^{\ast }$. In a second appendix we
present in greater detail than in \cite{7849} and on two examples, an
adaptation of the procedure of O'Dwyer and Osborn \cite{6228} for
determining an EMRO for a given Wilsonian ERGE.

\section{Summary of previous episodes\label{Sec:Sum}}

In order to well expose the issue discussed in this article it is useful to
first present a brief summary of previous episodes of the ERGE history.
Several modern reviews or lectures on the subject are available in the
literature \cite{4595,4700,6699}\cite{3993}{--}\cite{8089}.

\subsection{Notations\label{Nota}}

Let us first introduce some of our notation conventions.

The scale-de\-pen\-dent action $S_{\Lambda }\left[ \phi \right] $ and the
scale-de\-pen\-dent effective action $\Gamma _{\Lambda }\left[ M\right] $ will
be noted respectively $S\left[ \tilde{\phi},t\right] $ and $\Gamma \left[ 
\tilde{M},t\right] $ in which: 
\begin{equation}
t=-\ln \left( \Lambda /\mu _{0}\right) \,,  \label{eq:RGtime}
\end{equation}%
is the (dimensionless) RG-time, $\mu _{0}$ being an arbitrary momentum-scale
of reference larger than $\Lambda $ and which is different from the overall
cutoff $\Lambda _{0}$ (that will finally be sent to infinity)

The rescaling step of the RG\ procedure (RG-step 2) may be accounted for by
expressing the dimensions via factorized powers of $\Lambda $. In the
following we denote dimensionless quantities by letter with an upper tilde,
such as, for example:%
\begin{eqnarray}
q &=&\Lambda \tilde{q}\,,  \label{eq:dimq} \\
\phi \left( x\right) &=&\Lambda ^{d_{\phi }^{\left( c\right) }}\tilde{\phi}%
\left( \tilde{x}\right) \,,  \label{eq:dimphicla}
\end{eqnarray}%
in which $d_{\phi }^{\left( c\right) }$ is the classical dimension of the
field ($\phi \left( x\right) $ or $M\left( x\right) $). In order to make
contact with the historic first version of the ERGE \cite{440} we assume the
unusual value (see footnote \ref{foot1}):%
\begin{equation}
d_{\phi }^{\left( c\right) }=\frac{d}{2}-n_{0}\,,  \label{eq:dc}
\end{equation}%
where $d$ is the spatial dimension and $n_{0}$ will take on the value $0$ or $1$ according to the choice of
the cutoff function [see eq. (\ref{eq:defP})]. We shall show that Wilson's
choice:

\begin{equation}
d_{\phi }^{\left( cw\right) }=\frac{d}{2}\,,  \label{eq:dcwil}
\end{equation}%
corresponding to $n_{0}=0$, is linked to the possibility of getting rid of
the regularization procedure via field-redefinitions.

We will use also the following scale-de\-pen\-dent \textquotedblleft
anomalous\textquotedblright\ dimensions:%
\begin{eqnarray}
d_{\phi }^{\left( \pm \right) }\left( t\right) &=&\frac{d}{2}\pm \varpi
_{0}\left( t\right) \,,  \label{eq:dphi+-} \\
\varpi _{n_{0}}\left( t\right) &=&1-n_{0}-\frac{\eta \left( t\right) }{2}\,,
\label{eq:cdet}
\end{eqnarray}%
in which the function $\eta \left( t\right) $ is determined so as to keep
constant (i.e. independent of $t$) the coefficient of one term\footnote{%
This step is required due to the reparametrization invariance that implies
that the flow of one coefficient of $S$ is redundant. To eliminate this
redundancy, and to set the momentum scales, one usually keeps constant the
coefficient of the kinetic term along the flow of $S$.} of $S$. At a fixed
point, $\eta \left( t\right) $ is a constant $\eta ^{\ast }$ called the
anomalous dimension of the field (see appendix \ref{defZ}).

Sometimes, when no confusion may arise with the usual actions, we lightly
refer to the scale-de\-pen\-dent actions as $S$ and $\Gamma $. When needed, we
still refer to dimensioned quantities as previously ($\Lambda ,\phi ,x,q,M,$
etc.).

We also use sometimes the following writing conventions:

\begin{itemize}
\item $A_{q}$, $B\left( q^{2}\right) $ are Fourier transformed of
respectively $A\left( x\right) $, $B\left( x,y\right) $ when the usual
invariances by translation and rotation in space are assumed.

\item $A\cdot C$ stands for $\int d^{d}xA\left( x\right) C\left( x\right) $
or $\int_{q}A_{q}C_{-q}$

\item $A\cdot B\cdot C$ stands for $\int d^{d}xd^{d}yA\left( x\right)
B\left( x,y\right) C\left( y\right) $ or $\int_{q}A_{q}B\left( q^{2}\right)
C_{-q}$.
\end{itemize}

For the sake of an easy understanding of the present article it is important
to recall that the field-renormalization function $Z_{3}\left( \ell \right) $
is usually utilized to define the renormalized field $\phi _{R}$ as follows%
\footnote{%
Wetterich's field-renormalization fonction $Z_{\Lambda }$ is the inverse of $%
Z_{3}$ (see section \ref{Implem}) but this is a matter of convention
provided that the definition of the anomalous dimension $\eta \left(
t\right) $ be unchanged (see appendix \ref{defZ}).}:%
\begin{equation*}
\phi =\sqrt{Z_{3}\left( \ell \right) }\,\phi _{R}\,,\quad \mathrm{with}\quad
\ell <1\,.
\end{equation*}

Most often, to simplify the writing, we shall drop the subscript
\textquotedblleft $R$\textquotedblright\ when no confusion may arise.

So defined, $Z_{3}\left( \ell \right) $ is linked to the anomalous dimension
of the field $\eta \left( t\right) $ as (see appendix \ref{defZ}):%
\begin{equation}
\ell \frac{d}{d\ell }Z_{3}\left( \ell \right) =-2\,\varpi _{n_{0}}\left(
t\right) Z_{3}\left( \ell \right) \,,  \label{eq:Z3}
\end{equation}%
in which $n_{0}$ is defined in eq. (\ref{eq:defP}) in order to account for
different versions of ERGE. In (\ref{eq:Z3}) we have intentionally
distinguished $\ell $ from $e^{-t}$ with $t$ defined by (\ref{eq:RGtime});
this distinction has some importance as discussed in section \ref{ana}.

\subsection{The historic Wilson ERGE\label{WilERGE}}

The historic first ERGE \cite{440} has been set up directly for the
scale-de\-pen\-dent full action $S\left[ \tilde{\phi};t\right] $, including
rescaling and field-renormalization, it may be written under the following
form \cite{7849}:

\begin{eqnarray}
\dot{S}\left[ \tilde{\phi};t\right] &=&\int_{\tilde{q}}2\tilde{q}^{2}\left( 
\frac{\delta ^{2}S}{\delta \tilde{\phi}_{\tilde{q}}\delta \tilde{\phi}_{-%
\tilde{q}}}-\frac{\delta S}{\delta \tilde{\phi}_{\tilde{q}}}\frac{\delta S}{%
\delta \tilde{\phi}_{-\tilde{q}}}+\tilde{\phi}_{\tilde{q}}\frac{\delta S}{%
\delta \tilde{\phi}_{\tilde{q}}}\right)  \notag \\
&&+\mathcal{G}_{\mathrm{dil}}\left( S,\tilde{\phi},d_{\phi }^{\left(
cw\right) }\right) +\varpi _{0}\left( t\right) \mathcal{O}_{w}\left( S,%
\tilde{\phi}\right) \,,  \label{eq:ERGEWilHist}
\end{eqnarray}%
in which $\dot{S}\left[ \tilde{\phi};t\right] $ stands for $\left. dS\left[ 
\tilde{\phi};t\right] /dt\right\vert _{\tilde{\phi}}$.

The first line of (\ref{eq:ERGEWilHist}) corresponds to a realization of the
RG-step 1 (decimation) by means of an \textquotedblleft incomplete
integration\textquotedblright .

The second line shows two parts:

\begin{enumerate}
\item the rescaling part (RG-step 2): 
\begin{equation}
\mathcal{G}_{\mathrm{dil}}\left( S,\tilde{\phi},d_{\phi }\right) =\int_{%
\tilde{q}}\left[ \left( d-d_{\phi }\right) \,\tilde{\phi}_{\tilde{q}}+%
\mathbf{\tilde{q}}\cdot \frac{\partial }{\partial \mathbf{\tilde{q}}}\tilde{%
\phi}_{\tilde{q}}\right] \,\frac{\delta S}{\delta \tilde{\phi}_{\tilde{q}}}%
\,,  \label{eq:Gdil}
\end{equation}%
in which $d_{\phi }=d_{\phi }^{\left( cw\right) }$ is the classical
dimension of the field, it is given by (\ref{eq:dcwil}) --i.e. (\ref{eq:dc})
with\footnote{%
Thecnically, the form of the first line of (\ref{eq:ERGEWilHist}) implies
that $n_{0}=0$ \cite{7289}.} $n_{0}=0$.

\item the field-renormalization part (RG-step 3) with:%
\begin{equation}
\mathcal{O}_{w}\left( S,\tilde{\phi}\right) =\int_{\tilde{q}}\left( \frac{%
\delta ^{2}S}{\delta \tilde{\phi}_{\tilde{q}}\delta \tilde{\phi}_{-\tilde{q}}%
}-\frac{\delta S}{\delta \tilde{\phi}_{\tilde{q}}}\frac{\delta S}{\delta 
\tilde{\phi}_{-\tilde{q}}}+\tilde{\phi}_{\tilde{q}}\frac{\delta S}{\delta 
\tilde{\phi}_{\tilde{q}}}-1\right) \,.  \label{eq:Ow1}
\end{equation}
\end{enumerate}

By separating the contribution proportionnal to $\mathcal{O}_{w}\left( S,%
\tilde{\phi}\right) $ in (\ref{eq:ERGEWilHist}), we want to highlight the
following ideas that, presumably, are at the source of Wilson's particular
implemention of RG-step 3.

At a fixed point $S^{\ast }$ of (\ref{eq:ERGEWilHist}), $\eta \left(
t\right) $ takes on the constant value $\eta ^{\ast }$ and $\mathcal{O}%
_{w}\left( S^{\ast },\tilde{\phi}\right) $ is an EMRO responsible for an
(infinitesimal) generation of a line of equivalent fixed points under the
change $\tilde{\phi}\rightarrow \,\tilde{\phi}\left( 1+\delta a\right) $, $%
\delta a$ being a pure (infinitesimal) constant independent of $t$. Indeed,
as defined in (\ref{eq:Ow1}), $\mathcal{O}_{w}\left( S,\tilde{\phi}\right) $
is the infinitesimal version of the transformation\footnote{%
Here $\kappa $ may depend on $t$.} $U_{\kappa }$, introduced by Bell and
Wilson \cite{4421} (see also \cite{4405}) that commutes with the RG
procedure and is associated to a change of $\tilde{\phi}$ into $\tilde{\phi}%
^{\prime }=\kappa \tilde{\phi}$:%
\begin{equation}
U_{\kappa }\exp \left( -S\left[ \tilde{\phi}\right] \right) \propto \int 
\mathcal{D}\tilde{\phi}\exp \left\{ -\frac{1}{2\left( 1-\kappa ^{2}\right) }%
\int_{q}\left\vert \kappa \tilde{\phi}_{q}-\tilde{\phi}_{q}^{\prime
}\right\vert ^{2}-S\left[ \tilde{\phi}\right] \right\} \,.  \label{eq:U}
\end{equation}%
For $\kappa $ infinitesimally close to unity, the properties of the Gaussian
integral yields the expression (\ref{eq:Ow1}).

As any redundant operator $\mathcal{O}\left( S,\tilde{\phi}\right) $
expresses also under the following form \cite{4011,2835}:%
\begin{equation}
\mathcal{O}\left( S,\tilde{\phi}\right) =\int_{\tilde{q}}\left[ \psi _{%
\tilde{q}}\left( \tilde{\phi},S\right) \frac{\delta S}{\delta \tilde{\phi}_{%
\tilde{q}}}-\frac{\delta \psi _{\tilde{q}}\left( \tilde{\phi},S\right) }{%
\delta \tilde{\phi}_{\tilde{q}}}\right] \,,  \label{eq:Ogenw}
\end{equation}%
then for $\mathcal{O}_{w}\left( S,\tilde{\phi}\right) $ given by (\ref%
{eq:Ow1}) we have: 
\begin{equation}
\psi _{\tilde{q}}^{\left( w\right) }\left( \tilde{\phi},S\right) =\tilde{\phi%
}_{\tilde{q}}-\frac{\delta S}{\delta \tilde{\phi}_{-\tilde{q}}}\,.
\label{eq:PsiOwil}
\end{equation}

Contrary to what one might think, the expression of an EMRO is not
necessarily linked to $U_{\kappa }$ because that expression depends on
conventions and on the way both the field-renormalization and the RG-step 1
are realized (e.g., see appendix \ref{ObtainEMRO}).

In the following we shall call Wilsonian EMRO an EMRO formed by a linear
combination of $\tilde{\phi}_{\tilde{q}}$ and $\frac{\delta S}{\delta \tilde{%
\phi}_{-\tilde{q}}}$ such as:%
\begin{equation}
\psi _{\tilde{q}}^{\left( wil\right) }\left( \tilde{\phi},S\right) =\bar{a}%
\left( \tilde{q}^{2}\right) \tilde{\phi}_{\tilde{q}}+\bar{b}\left( \tilde{q}%
^{2}\right) \frac{\delta S}{\delta \tilde{\phi}_{-\tilde{q}}} \,.
\label{eq:PsiWil}
\end{equation}

\begin{remark}
It is to be noticed that, in the procedure described in \cite{440}, there is no overall UV-cutoff $\Lambda _{0}$
explicitly considered. The
procedure is so that the fields with large $q$ are merely
more integrated than those with small $q$ (without having to refer to any
explicit existence of a cutoff $\Lambda _{0}$ that would distort the action $S$). Only an initial RG-time $t=0$ is defined at
which the \textquotedblleft incomplete integration\textquotedblright\ is
started. It is like the introduction of the formal scale $\mu_0$ in the definition (\ref{eq:RGtime}) of $t$, which never appears
 explicitly in the ERGE.
Any advised user of ERGE knows that the
\textquotedblleft bare action\textquotedblright\ is merely a simple form of an action
chosen at will at $t=0$ as initial condition to the ERGE without having to specify any regularization process. Thus, to use a common language, in the Wilson ERGE (\ref{eq:ERGEWilHist}), the limit $\Lambda _{0}\rightarrow
\infty $ is implicitly performed .\label{UVWil}
\end{remark}

\begin{remark}
In conjunction with the preceding remark and for later use (see section \ref{ana}), it is interesting to recall here (in
other terms) a remark of Bell and Wilson \cite{4421}: this way of
renormalizing the field (using $U_{\kappa }$ and thus an EMRO), affects
\textquotedblleft every old\textquotedblright\ fields $\phi $ in contrast
with the reduction of variables effected by RG-step 1 that affects only a part of them.\label{Bell}
\end{remark}

\begin{remark}
The integral over $\tilde{q}$ of the historic first ERGE (\ref%
{eq:ERGEWilHist}) is not necessarily well defined for large $\tilde{q}$ so
that a redefinition of the field $\tilde{\phi}_{\tilde{q}}\rightarrow
g\left( \tilde{q}^{2}\right) \tilde{\phi}_{\tilde{q}}$ with $g\left( \tilde{q%
}^{2}\right) \rightarrow \infty $ sufficiently rapidly (and $g\left(0\right)=1$ to preserve the quasi-locality of the action), could be required
prior to explicit calculations, e.g., as in the framework of the derivative
expansion \cite{212}. This field redefinition is allowed owing to the
property of reparametrization invariance and could be added to the list of
RG-steps as an optional fourth step that may be implemented at will.%
\label{Rem:Wil0}
\end{remark}

\subsection{The Polchinski ERGE\label{PolERGE}}

Based on an explicit reference to an arbitrary smooth-UV-cutoff function,
Polchinski \cite{354} has constructed a RG flow equation for a truncation of 
$S_{\Lambda }\left[ \phi \right] $, denoted here $S_{\mathrm{int},\Lambda }%
\left[ \phi \right] $ and defined as follows:%
\begin{equation}
S_{\Lambda }\left[ \phi \right] =\frac{1}{2}\phi \cdot P^{-1}\cdot \phi +S_{%
\mathrm{int},\Lambda }\left[ \phi \right] \,,  \label{eq:Sint}
\end{equation}

In (\ref{eq:Sint}), a quadratic part has been evidenced in order to make
explicit (though arbitrary) the UV cutoff function $P\left( q^{2},\Lambda
\right) $ that, for the sake of comparison with the Wilson procedure
recalled in section \ref{WilERGE}, we choose to write under the form:%
\begin{equation}
P\left( q^{2},\Lambda \right) =\frac{K\left( \tilde{q}^{2}\right) }{\left(
q^{2}\right) ^{n_{0}}}\,,  \label{eq:defP}
\end{equation}%
in which $n_{0}$ may take on the value $0$ or $1$, and $K$ is dimensionless
with the assumed property that\footnote{%
Polchinski has in fact chosen a dimensioned form of $P$ namely: $P\left(
q^{2},\Lambda \right) =K\left( \frac{q^{2}}{\Lambda ^{2}}\right) /\left(
q^{2}+m^{2}\right) $ with $K$ dimensionless and, thus, with a classical
dimension $d_{\phi }^{\left( c\right) }=\left( d-2\right) /2$ for the field,
i.e. with $n_{0}=1$. In that case: $\lim_{\Lambda \rightarrow \infty
}P\left( q^{2},\Lambda \right) \propto \frac{1}{q^{2}+m^{2}}$. The idea
behind this choice is reminiscent of the old perturbative version of the RG,
in which the cutoff $\Lambda $ is seen as an artificial parameter that
modifies the propagator only temporarily. In the perturbative view, the
scale $\Lambda $ is introduced \textquotedblleft by hand\textquotedblright\
(and this arbitrariness is intended to be finally \textquotedblleft washed
out\textquotedblright\ in the final theory). This is in contrast to Wilson's
view in which $\Lambda $ is seen as an integral parameter of the RG\ theory.%
\label{foot1}} 
\begin{equation}
\lim_{\Lambda \rightarrow \infty }K\left( \tilde{q}^{2}\right) =\mathrm{%
const.}  \label{eq:limP}
\end{equation}%
So defined, $P$ has the dimension $-2n_{0}$.

\subsubsection{Decimation}

RG-step 1, expressed as a variation of $S_{\mathrm{int},\Lambda }$ under an
infinitesimal change of $\Lambda $, leads to what is usually named the
Polchinski ERGE \cite{354} (discussing the way it is obtained is out of the
scope of the present article, readers interesting by that issue may look
also at \cite{409,4442}):

\begin{equation}
\left. -\Lambda \frac{\partial }{\partial \Lambda }S_{\mathrm{int},\Lambda
}\right\vert _{\phi }=\frac{1}{2}\int_{q}\left. \Lambda \frac{\partial }{%
\partial \Lambda }P\left( q^{2},\Lambda \right) \right\vert _{q}\left[ \frac{%
\delta ^{2}S_{\mathrm{int},\Lambda }}{\delta \phi _{q}\delta \phi _{-q}}-%
\frac{\delta S_{\mathrm{int},\Lambda }}{\delta \phi _{q}}\frac{\delta S_{%
\mathrm{int},\Lambda }}{\delta \phi _{-q}}\right] \,.  \label{eq:PolchBrute}
\end{equation}

Coming back to the full action $S_{\Lambda }\left[ \phi \right] $ via (\ref%
{eq:Sint}), (\ref{eq:PolchBrute}) gives (up to an additive constant term
which is currently neglected in the framework of field theory):%
\begin{equation*}
\left. -\Lambda \frac{\partial }{\partial \Lambda }S_{\Lambda }\right\vert
_{\phi }=\frac{1}{2}\int_{q}\left. \Lambda \frac{\partial }{\partial \Lambda 
}P\left( q^{2},\Lambda \right) \right\vert _{q}\left[ \frac{\delta
^{2}S_{\Lambda }}{\delta \phi _{q}\delta \phi _{-q}}-\frac{\delta S_{\Lambda
}}{\delta \phi _{q}}\frac{\delta S_{\Lambda }}{\delta \phi _{-q}}%
+2P^{-1}\phi _{q}\frac{\delta S_{\Lambda }}{\delta \phi _{q}}\right] \,.
\end{equation*}

\subsubsection{Rescaling}

The implementation of RG-step 2 (rescaling step) introduces dimensionless
quantities thus we use dimensionless field $\tilde{\phi}$ as defined in (\ref%
{eq:dimphicla}, \ref{eq:dc}) and we have:%
\begin{eqnarray*}
\frac{1}{2}\left. \Lambda \frac{\partial }{\partial \Lambda }P\left(
q^{2},\Lambda \right) \right\vert _{q} &=&\Lambda ^{-2n_{0}}\left[ -n_{0}%
\tilde{P}\left( \tilde{q}^{2}\right) -\tilde{q}^{2}\tilde{P}^{\prime }\left( 
\tilde{q}^{2}\right) \right]  \\
&=&-\Lambda ^{-2n_{0}}\left( \tilde{q}^{2}\right) ^{1-n_{0}}K^{\,\prime
}\left( \tilde{q}^{2}\right) \,,
\end{eqnarray*}%
where the prime denotes a derivative w.r.t. $\tilde{q}^{2}$. After
completing RG-step 2, but not yet RG-step 3 (field-renormalization), we get
for the full action:%
\begin{eqnarray}
\dot{S}\left[ \tilde{\phi},t\right] &=&-\int_{\tilde{q}}\left( \tilde{q}%
^{2}\right) ^{1-n_{0}}K^{\,\prime }\left( \tilde{q}^{2}\right) \left[ \frac{%
\delta ^{2}S}{\delta \tilde{\phi}_{\tilde{q}}\delta \tilde{\phi}_{-\tilde{q}}%
}-\frac{\delta S}{\delta \tilde{\phi}_{\tilde{q}}}\frac{\delta S}{\delta 
\tilde{\phi}_{-\tilde{q}}}+2\tilde{P}^{-1}\tilde{\phi}_{\tilde{q}}\frac{%
\delta S}{\delta \tilde{\phi}_{\tilde{q}}}\right]  \notag \\
&&+\mathcal{G}_{\mathrm{dil}}\left( S,\tilde{\phi},d_{\phi }^{\left(
c\right) }\right) \,.  \label{eq:PolchScaling}
\end{eqnarray}

It is interesting to perform also RG-step 2 on (\ref{eq:PolchBrute}). We get:%
\begin{eqnarray}
\dot{S}_{\mathrm{int}}\left[ \tilde{\phi},t\right] &=&-\int_{\tilde{q}%
}\left( \tilde{q}^{2}\right) ^{1-n_{0}}K^{\,\prime }\left( \tilde{q}%
^{2}\right) \left[ \frac{\delta ^{2}S_{\mathrm{int}}}{\delta \tilde{\phi}_{%
\tilde{q}}\delta \tilde{\phi}_{-\tilde{q}}}-\frac{\delta S_{\mathrm{int}}}{%
\delta \tilde{\phi}_{\tilde{q}}}\frac{\delta S_{\mathrm{int}}}{\delta \tilde{%
\phi}_{-\tilde{q}}}\right]  \notag \\
&&+\mathcal{G}_{\mathrm{dil}}\left( S_{\mathrm{int}},\tilde{\phi},d_{\phi
}^{\left( c\right) }\right) \,.  \label{eq:PolchDimLess}
\end{eqnarray}

It is a simple exercise to verify that this latter equation may be deduced
from (\ref{eq:PolchScaling}) via the dimensionless version of (\ref{eq:Sint}%
), namely:%
\begin{equation}
S\left[ \tilde{\phi},t\right] =\frac{1}{2}\tilde{\phi}\cdot \tilde{P}%
^{-1}\cdot \tilde{\phi}+S_{\mathrm{int}}\left[ \tilde{\phi},t\right] \,,
\label{eq:SintDimLess}
\end{equation}%
that induces the relation 
\begin{equation}
\dot{S}\left[ \tilde{\phi},t\right] =\dot{S}_{\mathrm{int}}\left[ \tilde{\phi%
},t\right] \,.  \label{eq:Spt=SintPt}
\end{equation}%
.

If, at this stage of the RG program, we perform the change $\tilde{\phi}%
\rightarrow \tilde{\phi}\sqrt{\tilde{P}}$ that implies $n_{0}=0$ (to
preserve quasi-locality) then [by virtue of (\ref{eq:Gdil}), with $d_{\phi
}^{\left( c\right) }\rightarrow d_{\phi }^{\left( cw\right) }$ given by (\ref%
{eq:dcwil}) and also $\tilde{P}=K$] equation (\ref{eq:PolchScaling}) reads:%
\begin{eqnarray}
\dot{S}\left[ \tilde{\phi},t\right] &=&-\int_{\tilde{q}}\tilde{q}^{2}\frac{%
\tilde{P}^{\,\prime }\left( \tilde{q}^{2}\right) }{\tilde{P}}\left[ \frac{%
\delta ^{2}S}{\delta \tilde{\phi}_{\tilde{q}}\delta \tilde{\phi}_{-\tilde{q}}%
}-\frac{\delta S}{\delta \tilde{\phi}_{\tilde{q}}}\frac{\delta S}{\delta 
\tilde{\phi}_{-\tilde{q}}}+2\tilde{\phi}_{\tilde{q}}\frac{\delta S}{\delta 
\tilde{\phi}_{\tilde{q}}}\right]  \notag \\
&&+\mathcal{G}_{\mathrm{dil}}\left( S,\tilde{\phi},d_{\phi }^{\left(
cw\right) }\right) +\int_{\tilde{q}}\tilde{q}^{2}\frac{\tilde{P}^{\,\prime
}\left( \tilde{q}^{2}\right) }{\tilde{P}}\tilde{\phi}_{\tilde{q}}\,\frac{%
\delta S}{\delta \tilde{\phi}_{\tilde{q}}}\,.  \label{eq:WilPochScaling}
\end{eqnarray}

\bigskip

Comparing with (\ref{eq:ERGEWilHist}) it is easy to verify that Wilson's
version corresponds to Polchinski's version (provided the field redefinition
into $\tilde{\phi}\sqrt{\tilde{P}\left( \tilde{q}^{2}\right) }$) with the
choice $\tilde{P}\left( \tilde{q}^{2}\right) =e^{-2\tilde{q}^{2}}$ \cite%
{3357}. One could already notice the difference of spirit induced by this
field redefinition: the cutoff function has finally been removed from the
action so that $\Lambda $ appears to be a pure scale $\Lambda =e^{-t}\mu
_{0} $ (see remark \ref{UVWil}).

It is also interesting to notice that in the Polchinski approach, the
running cutoff $\Lambda $ itself is an overall UV-cutoff $\Lambda _{0}$.
Consequently, at this stage, it is not obvious to see how it is possible to
reproduce in a Polchinski ERGE, the limit $\Lambda _{0}\rightarrow \infty $
at fixed $\Lambda $ that is commonly performed and discussed in the case of
a Wetterichian ERGE (see section \ref{WettERGE}). That important issue is
discussed in section \ref{Limite}.

The correspondence with the Wilson version established just above is not
complete since it has been obtained in absence of field-renormalization.
Commonly, this latter RG-step is completed following the procedure of Ball
et al \cite{3491}, who did not introduce $\eta \left( t\right) $ the same
way as Wilson did to get (\ref{eq:ERGEWilHist}).

\subsubsection{Field-renormalization: the \textquotedblleft
modified\textquotedblright\ Polchinski ERGE \label{modif}}

The implementation of RG-step 3 in the flow equation for\ $S_{\mathrm{int}}%
\left[ \tilde{\phi},t\right] $ can be regularly realized by replacing in (%
\ref{eq:PolchDimLess}) $d_{\phi }^{\left( c\right) }$ by $d_{\phi }^{\left(
-\right) }$ as defined in eq. (\ref{eq:dphi+-}) to get:%
\begin{eqnarray}
\dot{S}_{\mathrm{int}}\left[ \tilde{\phi},t\right]  &=&-\int_{\tilde{q}%
}\left( \tilde{q}^{2}\right) ^{1-n_{0}}K^{\,\prime }\left( \tilde{q}%
^{2}\right) \left[ \frac{\delta ^{2}S_{\mathrm{int}}}{\delta \tilde{\phi}_{%
\tilde{q}}\delta \tilde{\phi}_{-\tilde{q}}}-\frac{\delta S_{\mathrm{int}}}{%
\delta \tilde{\phi}_{\tilde{q}}}\frac{\delta S_{\mathrm{int}}}{\delta \tilde{%
\phi}_{-\tilde{q}}}\right]   \notag \\
&&+\mathcal{G}_{\mathrm{dil}}\left( S_{\mathrm{int}},\tilde{\phi},d_{\phi
}^{\left( -\right) }\right) \,.  \label{eq:PolchLin}
\end{eqnarray}

Indeed, this procedure amounts to expressing on $S_{\mathrm{int}}$ the
regular infinitesimal field-renormalization associated to an infinitesimal
realization of RG-step 1. This way of introducing $\eta \left( t\right) $,
the common way, is said \textquotedblleft linear\textquotedblright\ in \cite%
{7849} by reference to the fact that it contributes linearly in $S_{\mathrm{%
int}}$ (or $S$) in the flow equation, in contrast to the Wilson procedure
which is \textquotedblleft non-linear\textquotedblright\ in this respect.

In practice\footnote{%
The reasoning of Ball et al is different but it is more convenient, and
equivalent, to present it that way.}, instead of considering $S_{\mathrm{int}%
}$, Ball et al \cite{3491} have chosen to perform the above change $d_{\phi
}^{\left( c\right) }$by $d_{\phi }^{\left( -\right) }$ within the flow
equation of the \textsl{full action} $S\left[ \tilde{\phi},t\right] $ (\ref%
{eq:PolchScaling}) which then reads: 
\begin{eqnarray}
\dot{S}\left[ \tilde{\phi},t\right] &=&-\int_{\tilde{q}}\left( \tilde{q}%
^{2}\right) ^{1-n_{0}}K^{\,\prime }\left( \tilde{q}^{2}\right) \left[ \frac{%
\delta ^{2}S}{\delta \tilde{\phi}_{\tilde{q}}\delta \tilde{\phi}_{-\tilde{q}}%
}-\frac{\delta S}{\delta \tilde{\phi}_{\tilde{q}}}\frac{\delta S}{\delta 
\tilde{\phi}_{-\tilde{q}}}+2\tilde{P}^{-1}\tilde{\phi}_{\tilde{q}}\frac{%
\delta S}{\delta \tilde{\phi}_{\tilde{q}}}\right]  \notag \\
&&+\mathcal{G}_{\mathrm{dil}}\left( S,\tilde{\phi},d_{\phi }^{\left(
-\right) }\right) \,.  \label{eq:WilModif}
\end{eqnarray}

(Because the implementation of RG-step 3 is different from the original
Wilson way that yields (\ref{eq:ERGEWilHist}), we refer to (\ref{eq:WilModif}%
)\ as the \textquotedblleft modified\textquotedblright\ Wilson ERGE extended
to an arbitrary cutoff.)

Assuming\footnote{%
This assumption is not obviously justified, see remark \ref{rem1}.} that the
effect of the infinitesimal implementation of \ RG-step 3 does not affect
eq. (\ref{eq:SintDimLess}), and, consequently, that the condition (\ref%
{eq:Spt=SintPt}) is satisfied (see remark \ref{rem1} below), one finally
gets the flow equation of Ball et al for $S_{\mathrm{int}}$: 
\begin{eqnarray}
\dot{S}_{\mathrm{int}}\left[ \tilde{\phi},t\right]  &=&-\int_{\tilde{q}%
}\left( \tilde{q}^{2}\right) ^{1-n_{0}}K^{\,\prime }\left( \tilde{q}%
^{2}\right) \left[ \frac{\delta ^{2}S_{\mathrm{int}}}{\delta \tilde{\phi}_{%
\tilde{q}}\delta \tilde{\phi}_{-\tilde{q}}}-\frac{\delta S_{\mathrm{int}}}{%
\delta \tilde{\phi}_{\tilde{q}}}\frac{\delta S_{\mathrm{int}}}{\delta \tilde{%
\phi}_{-\tilde{q}}}\right]   \notag \\
&&+\mathcal{G}_{\mathrm{dil}}\left( S_{\mathrm{int}},\tilde{\phi},d_{\phi
}^{\left( -\right) }\right) -\varpi _{n_{_{0}}}\left( t\right) \,\tilde{\phi}%
\cdot \tilde{P}^{-1}\cdot \tilde{\phi}\,.  \label{eq:PolchModif}
\end{eqnarray}%
This equation has been called the \textquotedblleft
modified\textquotedblright\ Polchinski flow equation \cite{7289,7205,6750}
because it involves an unusual\ quadratic term proportionnal to\footnote{%
In the version of Ball et al \cite{3491} this term is proportionnal to $%
\varpi _{1}\left( t\right) =-\eta \left( t\right) /2$ due to the choice of a
dimensioned cutoff function $P\left( q^{2},\Lambda \right) $ implying $%
n_{0}=1$, see footnote \ref{foot1}.} $\varpi _{n_{_{0}}}\left( t\right) $ as
defined in (\ref{eq:cdet}).

\begin{remark}
The additional Gaussian term that characterizes the \textquotedblleft
modified\textquotedblright\ flow equation [last\ term of (\ref{eq:PolchModif}%
)] is due to the fact that the modification of the quadratic term of (\ref%
{eq:Sint}) induced by the field-renormalisation step is not compensated by a
scale-de\-pen\-dent counter-part within the cutoff function $P$. Its presence in
the RG-flow equation is actually due to the arbitrary condition (\ref%
{eq:Spt=SintPt}). If, instead, we impose the more regular condition:%
\begin{equation}
\dot{S}=\dot{S}_{\mathrm{int}}+\varpi_{n_0}\left( t\right) \,\tilde{\phi}\cdot \tilde{P}%
^{-1}\cdot \tilde{\phi} \,,  \label{eq:SptTranslated}
\end{equation}%
then the RG-flow equation for $S_{\mathrm{int}}$ reduces merely to (\ref%
{eq:PolchLin}) which is also a well acceptable flow equation.\label{rem1}
\end{remark}

\subsubsection{Complicated expression of an EMRO for the \textquotedblleft
modified\textquotedblright\ Wilson ERGE\label{EMROWilMod}}

Beyond the fact that the original Wilson ERGE is obviously not recovered
with the common (linear) way of completing the field-renormalization (what
is a priori nothing to be ashamed of), there are important consequences.

In particular, the expression of a Wilsonian EMRO [see eq. (\ref{eq:PsiWil}%
)] --responsible for the existence of lines of equivalent fixed points--
takes on a complicated form with the common procedure \cite{6228} whereas
this form is simple with the Wilson procedure (see also appendix \ref%
{ObtainEMRO}).

In order to get the expression of the Wilsonian EMRO associated to the
\textquotedblleft modified\textquotedblright\ Wilsonian ERGE (\ref%
{eq:WilModif}), O'Dwyer and Osborn \cite{6228} have looked for an EMRO,
corresponding to $\psi _{\tilde{q}}^{\left( wil\right) }\left( \tilde{\phi}%
,S\right) $ as given by (\ref{eq:PsiWil}).

In the case where the cutoff function is given by (\ref{eq:defP}) with $%
n_{0}=1$, they have obtained\footnote{%
The functions $\bar{a}$ and $\bar{b}$ are the functions $a$ and $b$ of \cite{6228} multiplied by $K\left( x\right) $.}:%
\begin{equation}
\bar{a}\left( x\right) =1+x\bar{b}\left( x\right) ,\qquad \bar{b}\left(
x\right) =x^{\eta ^{\ast }/2-1}K^{2}\left( x\right) \int_{0}^{x}\,u^{-\eta
^{\ast }/2}\frac{K^{\prime }\left( u\right) }{\left[ K\left( u\right) \right]
^{2}}\,du\,.  \label{eq:abOsborn}
\end{equation}

This is in contrast with the Wilson procedure with which the chosen EMRO has
the simple form (\ref{eq:PsiOwil}) and, once \textquotedblleft \textsl{%
extended to an arbitrary cutoff function}\textquotedblright\ $\tilde{P}%
\left( \tilde{q}^{2}\right) $, reads (provided that $n_{0}=0$) \cite{7849}:%
\begin{equation*}
\psi _{q,\mathcal{O}}=\tilde{\phi}_{q}-\tilde{P}\left( \tilde{q}^{2}\right) 
\frac{\delta S^{\ast }}{\delta \tilde{\phi}_{-q}}\,.
\end{equation*}

In appendix \ref{ObtainEMRO} we determine a similar simple expression of an
EMRO in the Polchinski case $n_{0}=1$ when RG-step 3 is not linearly
implemented.

Similarly to the expression of the EMRO given by (\ref{eq:PsiWil}, \ref%
{eq:abOsborn}), the Legendre transformation that links $S_{\mathrm{int}}$
and $\Gamma _{\mathrm{int}}$ and their respective RG-flow equations has been
found to be extremely complicated \cite{7289,7205,6750} when the
field-renormalization step is implemented the common way. From here it is
but a short step to believing that with a Wilson-like non-linear
implementation of the field-renormalization step (i.e., not linearly and via
an EMRO), the Legendre transformation would take on a simpler form. Let us
first recall how the RG-flow equation for $\Gamma _{\mathrm{int}}$ has been
commonly treated.

\subsection{The Wetterich ERGE and the Legendre transformation\label%
{WettERGE}}

We call Wetterich's flow equation, the ERGE for the effective average action 
$\Gamma _{\mathrm{int}}$ (for a review, see \cite{4700}).

\subsubsection{Decimation}

There are several ways to get this kind of flow equation \cite%
{4374,4436,2520,4293}. For our purposes, the most convenient way is to
derive it from the Polchinki ERGE via a Legendre transformation \cite%
{4436,2520}. Let us summarize sketchily the main steps of that derivation
(for more details see, e.g. \cite{2520}).

We start with the scale-de\-pen\-dent action $S_{\Lambda }$ as Polchinski did
with (\ref{eq:Sint}) but, for convenience, we consider an initial (overall)
arbitrary cutoff $\Lambda _{0}$ that is greater than the \textquotedblleft
running\textquotedblright\ cutoff $\Lambda $ of the preceding sections. This
overall momentum scale is implemented by an arbitrary UV cutoff function $%
\Delta _{0}\left( q^{2},\Lambda _{0}\right) $, similar to $P$ in (\ref%
{eq:Sint}):%
\begin{equation}
S_{\Lambda _{0}}\left[ \phi \right] =\frac{1}{2}\phi \cdot \Delta
_{0}^{-1}\cdot \phi +S_{\mathrm{int},\Lambda _{0}}\left[ \phi \right] \,.
\label{eq:Sint0}
\end{equation}

Based on the property that the partition function associated to $S_{\Lambda
_{0}}\left[ \phi \right] $ can be rewritten in terms of two quadratic parts
and two fields such as $\phi =\phi _{1}+\phi _{2}$, we write:%
\begin{equation*}
S_{\Lambda _{0}}\left[ \phi \right] =\frac{1}{2}\phi _{1}\cdot \Delta
_{1}^{-1}\cdot \phi _{1}+\frac{1}{2}\phi _{2}\cdot \Delta _{2}^{-1}\cdot
\phi _{2}+S_{\mathrm{int},\Lambda _{0}}\left[ \phi _{1}+\phi _{2}\right] \,,
\end{equation*}%
where $\Delta _{1}\left( q^{2};\Lambda \right) $ is an UV cutoff function
associated to $\Lambda $ and $\Delta _{2}\left( q^{2};\Lambda ,\Lambda
_{0}\right) $ an IR cutoff function defined as: 
\begin{equation}
\Delta _{2}\left( q^{2};\Lambda ,\Lambda _{0}\right) =\Delta _{0}\left(
q^{2},\Lambda _{0}\right) -\Delta _{1}\left( q^{2},\Lambda \right) \,.
\label{eq:Delta2}
\end{equation}%
Integrating out partially over $\phi _{2}$ generates $S_{\mathrm{int}%
,\Lambda }\left[ \phi _{1}\right] _{\Lambda }$ which, using a property of
the Gaussian integral, expresses as follows:%
\begin{equation}
\exp \left( -S_{\mathrm{int},\Lambda }\left[ \phi _{1}\right] \right) =\exp
\left( \frac{1}{2}\frac{\delta }{\delta \phi _{1}}\cdot \Delta _{2}\cdot 
\frac{\delta }{\delta \phi _{1}}\right) \exp \left\{ -S_{\mathrm{int}%
,\Lambda _{0}}\left[ \phi _{1}\right] \right\} \,.  \label{eq:IntegForm}
\end{equation}

A derivative w.r.t. $\Lambda $ of (\ref{eq:IntegForm}) gives:

\begin{equation}
\left. \Lambda \frac{\partial S_{\mathrm{int},\Lambda }}{\partial \Lambda }%
\right\vert _{\phi _{1}}=\frac{1}{2}\int_{q}\left. \Lambda \frac{\partial
\Delta _{2}\left( q^{2};\Lambda ,\Lambda _{0}\right) }{\partial \Lambda }%
\right\vert _{q}\left[ \frac{\delta ^{2}S_{\mathrm{int},\Lambda }}{\delta
\phi _{1},_{q}\delta \phi _{1,-q}}-\frac{\delta S_{\mathrm{int},\Lambda }}{%
\delta \phi _{1},_{q}}\frac{\delta S_{\mathrm{int},\Lambda }}{\delta \phi
_{1,-q}}\right] \,,  \label{eq:PolchBrute2}
\end{equation}%
which is the Polchinski flow equation (\ref{eq:PolchBrute}) provided that 
\cite{2599}:

\begin{equation}
\Lambda \frac{\partial }{\partial \Lambda }\Delta _{2}=-\Lambda \frac{%
\partial }{\partial \Lambda }P\,,  \label{eq:Delta2P}
\end{equation}%
which, from (\ref{eq:Delta2}) is satisfied if 
\begin{equation}
P\left( q^{2},\Lambda \right) =\Delta _{1}\left( q^{2},\Lambda \right) \,.
\label{eq:Delta1P}
\end{equation}

Though we may define a scale-de\-pen\-dent \textit{full} action $S_{\Lambda }%
\left[ \phi _{1}\right] $ as:%
\begin{equation}
S_{\Lambda }\left[ \phi _{1}\right] =\frac{1}{2}\phi _{1}\cdot \Delta
_{1}^{-1}\cdot \phi _{1}+S_{\mathrm{int},\Lambda }\left[ \phi _{1}\right] \,,
\label{eq:Sint1}
\end{equation}%
the effective Legendre transformation of interest involves $S_{\mathrm{int}%
,\Lambda }\left[ \phi _{1}\right] _{\Lambda }$ and a truncated
scale-de\-pen\-dent effective action $\Gamma _{\mathrm{int},\Lambda }\left[ M_{1}%
\right] $ linked to the full scale-de\-pen\-dent effective action $\Gamma
_{\Lambda }\left[ M_{1}\right] $ as\footnote{%
The common use, inherited from the perturbative field theory, assumes a
cutoff function that must vanish when $\Lambda \rightarrow 0$ so that $%
\Gamma _{\mathrm{int,}\Lambda }\left[ M_{1}\right] \rightarrow \Gamma \left[
M_{1}\right] $ in this limit. Here we need not this condition because we
assume, following Wilson, that the quadratic term involving the cutoff
function is not an artificial part added by hand but an actual integral part
of the full scale-de\-pen\-dent effective action $\Gamma _{\Lambda }\left[ M_{1}%
\right] $.}:%
\begin{equation}
\Gamma _{\Lambda }\left[ M_{1}\right] =\frac{1}{2}M_{1}\cdot \Delta
_{2}^{-1}\cdot M_{1}+\Gamma _{\mathrm{int},\Lambda }\left[ M_{1}\right] \,.
\label{eq:FullGamma0}
\end{equation}%
The Legendre transformation thus reads \cite{2520}:%
\begin{eqnarray}
S_{\mathrm{int},\Lambda }\left[ \phi _{1}\right] &=&\frac{1}{2}\left(
M_{1}-\phi _{1}\right) \cdot \Delta _{2}^{-1}\cdot \left( M_{1}-\phi
_{1}\right) +\Gamma _{\mathrm{int},\Lambda }\left[ M_{1}\right] \,,
\label{eq0:TL1} \\
M_{1} &=&\phi _{1}-\Delta _{2}\cdot \frac{\delta }{\delta \phi _{1}}S_{%
\mathrm{int},\Lambda }\left[ \phi _{1}\right] \,.  \label{eq0:TL2}
\end{eqnarray}

Then, from (\ref{eq:PolchBrute2}) and using the properties:%
\begin{eqnarray}
\frac{\delta }{\delta \phi _{q}}S_{\mathrm{int},\Lambda }\left[ \phi _{1}%
\right] &=&\frac{\delta }{\delta M_{1,q}}\Gamma _{\mathrm{int},\Lambda }%
\left[ M_{1}\right] \,,  \label{eq0:Sp} \\
\frac{\delta ^{2}S_{\mathrm{int},\Lambda }\left[ \phi _{1}\right] }{\delta
\phi _{1},_{q}\delta \phi _{1,-q}} &=&\frac{\Gamma _{\mathrm{int},\Lambda
}^{\left( 2\right) }\left[ q;M_{1}\right] }{1+\Delta _{2}\left(
q^{2};\Lambda ,\Lambda _{0}\right) \Gamma _{\mathrm{int},\Lambda }^{\left(
2\right) }\left[ q;M_{1}\right] }\,,  \label{eq0:Spp}
\end{eqnarray}%
we obtain the RG equation of interest:%
\begin{equation}
\Lambda \frac{\partial }{\partial \Lambda }\Gamma _{\mathrm{int},\Lambda }%
\left[ M_{1}\right] =\frac{1}{2}\int_{q}\Lambda \frac{\partial \Delta
_{2}\left( q^{2};\Lambda ,\Lambda _{0}\right) }{\partial \Lambda }\frac{%
\Gamma _{\mathrm{int},\Lambda }^{\left( 2\right) }\left[ q;M_{1}\right] }{%
1+\Delta _{2}\left( q^{2};\Lambda ,\Lambda _{0}\right) \Gamma _{\mathrm{int}%
,\Lambda }^{\left( 2\right) }\left[ q;M_{1}\right] }\,.
\label{eq:WettBruteAvantLimite}
\end{equation}

\subsubsection{Limit of infinite overall cutoff}

To get the genuine Wetterich RG flow equation, the supplementary limit $%
\Lambda _{0}\rightarrow \infty $ (at fixed $\Lambda $) must be performed in (%
\ref{eq:WettBruteAvantLimite}) so as to finally obtain, up to a negligible
additive constant, the well-known expression:%
\begin{equation}
\Lambda \frac{\partial }{\partial \Lambda }\Gamma _{\mathrm{int},\Lambda }%
\left[ M_{1}\right] =\frac{1}{2}\int_{q}\Lambda \left. \frac{\partial
R\left( q^{2};\Lambda \right) }{\partial \Lambda }\right\vert _{q}\frac{1}{%
R\left( q^{2};\Lambda \right) +\Gamma _{\mathrm{int},\Lambda }^{\left(
2\right) }\left[ q;M_{1}\right] }\,,  \label{eq:WettBrute}
\end{equation}%
in which:%
\begin{equation}
R\left( q^{2};\Lambda \right) =\lim_{\Lambda _{0}\rightarrow \infty }\Delta
_{2}^{-1}\left( q^{2};\Lambda ,\Lambda _{0}\right) \,.
\label{eq:InfiniteCutoff}
\end{equation}

\subsubsection{Rescaling\label{rescaling}}

The rescaling step is easily accounted for by simple dimensional analysis
and leads to:%
\begin{equation}
\dot{\Gamma}_{\mathrm{int}}\left[ \tilde{M}_{1},t\right] =\int_{\tilde{q}}%
\frac{\left( -n_{0}\tilde{R}\left( \tilde{q}^{2}\right) +\tilde{q}^{2}\tilde{%
R}^{\prime }\left( \tilde{q}^{2}\right) \right) }{\tilde{R}\left( \tilde{q}%
^{2}\right) +\Gamma _{\mathrm{int}}^{\left( 2\right) }\left[ \tilde{q};%
\tilde{M}_{1}\right] }+\mathcal{G}_{\mathrm{dil}}\left( \Gamma _{\mathrm{int}%
},\tilde{M},d_{\phi }^{\left( c\right) }\right) \,,  \label{eq:Wett0}
\end{equation}%
in which $\mathcal{G}_{\mathrm{dil}}$ and $d_{\phi }^{\left( c\right) }$ are
defined respectively in (\ref{eq:Gdil}) and (\ref{eq:dc}) and the RG-time $t$
in (\ref{eq:RGtime}). This step is still compatible with the following
dimensionless form of (\ref{eq0:TL1}, \ref{eq0:TL2}):%
\begin{eqnarray}
S_{\mathrm{int}}\left[ \tilde{\phi}_{1},t\right] &=&\frac{1}{2}\left( \tilde{%
M}_{1}-\tilde{\phi}_{1}\right) \cdot \tilde{R}\cdot \left( \tilde{M}_{1}-%
\tilde{\phi}_{1}\right) +\Gamma _{\mathrm{int}}\left[ \tilde{M}_{1},t\right]
\,,  \label{eq0:TL1DimLess} \\
\tilde{M}_{1} &=&\tilde{\phi}_{1}-\tilde{R}^{-1}\cdot \frac{\delta }{\delta 
\tilde{\phi}_{1}}S_{\mathrm{int}}\left[ \tilde{\phi}_{1},t\right] \,,
\label{eq0:TL2DimLess}
\end{eqnarray}%
in the sense that, with the additional condition $\dot{S}_{\mathrm{int}}=%
\dot{\Gamma}_{\mathrm{int}}$, these relations enable us to recover (\ref%
{eq:PolchDimLess}) from (\ref{eq:Wett0}) --up to an additive constant--
provided that $\tilde{R}=-\left( \tilde{q}^{2}\right) ^{n_{0}}/\left(
K\left( \tilde{q}^{2}\right) -\mathrm{const}\right) $ which is compatible
with the dimensionless version of (\ref{eq:Delta1P}) and with the limit of
overall cutoff $\Lambda _{0}\rightarrow \infty $.

Describing the way RG-step 3 has been commonly implemented is a bit more
delicate.

\subsubsection{Field-renormalization\label{Implem}}

In principle, the field-renormalization step should be implemented purely
\textquotedblleft linearly\textquotedblright\ as done in section \ref{modif}%
. This would correspond to an infinitesimal realization of RG-step 3 in
response to the infinitesimal realization of RG-step 1. In practice it would
consist in changing $d_{\phi }^{\left( c\right) }$ by $d_{\phi }^{\left(
-\right) }$ [given by (\ref{eq:dphi+-})] within $\mathcal{G}_{\mathrm{dil}%
}\left( \Gamma _{\mathrm{int}},\tilde{M},d_{\phi }^{\left( c\right) }\right) 
$ appearing in (\ref{eq:Wett0}). It is easy to verify that, applying the
procedure of the preceding section to (\ref{eq:Wett0}) so modified, would
not enable us to recover (\ref{eq:PolchLin}) or (\ref{eq:PolchModif})
without modifying the Legendre transformation\footnote{%
For the same reason as one has \textquotedblleft modified\textquotedblright\
the Polchinski equation (see section \ref{modif}).}. This is because no
explicit scale-dependence has been assumed within the cutoff function of the
quadratic term in (\ref{eq0:TL1}) that could compensate the
field-renormalization effect.

It is a matter of fact that, instead of implementing it linearly (i.e.
infinitesimally), the current introduction of $\eta \left( t\right) $ within
the flow equation of the effective action is based on a cutoff function
involving a factorized scale-dependence that, precisely, compensates a
global field-renormalization conveniently added to the infinitesimal
implementation of the RG-steps. In the literature, we distinguish
essentially two (very close) ways to proceed, one is due to Morris \cite%
{3357} and the other to Wetterich (for a review, see \cite{4700}).

\paragraph{The Morris procedure}

It is based on an \textquotedblleft anomalous\textquotedblright\ dimensional
analysis that relies upon the fact that, at a fixed point or very close to
it, the effective dimension of the field is $d_{\phi }^{*\left( - \right) }$
[see (\ref{eq:dphi+-}) with $\eta \left( t\right) =\eta ^{\ast }$]. As a
consequence, to compensate this anomalous dimension introduced by hand, the
cutoff function $R$ (thus after the limit $\Lambda _{0}\rightarrow \infty $
has been performed) must display an anomalous dependence in the scale $%
\Lambda $ such as (in \cite{3357}, $R$ is noted $C^{-1}$):%
\begin{equation}
R\left( q^{2};\Lambda \right) =\Lambda ^{2-\eta ^{\ast }}\tilde{R}\left( 
\tilde{q}^{2}\right) \,.  \label{eq:RMorris}
\end{equation}

The derivative w.r.t. $\Lambda ,$ occuring in the right hand side of (\ref%
{eq:WettBrute}), induces within the flow equation a nonlinear (w.r.t. $%
\Gamma _{\mathrm{int}}$) contribution proportionnal to $\eta ^{\ast }$ in
addition to the usual linear contribution within $\mathcal{G}_{\mathrm{dil}%
}\left( \Gamma _{\mathrm{int}},\tilde{M},d_{\phi }^{\left( -\right) }\right) 
$. The ERGE for $\Gamma _{\mathrm{int}}$ reads thus \cite{3357}:%
\begin{equation}
\dot{\Gamma}_{\mathrm{int}}\left[ \tilde{M}_{1},t\right] =\frac{1}{2}\int_{%
\tilde{q}}\frac{\left[ \left( \eta ^{\ast }-2\right) \tilde{R}\left( \tilde{q%
}^{2}\right) +2\tilde{q}^{2}\tilde{R}^{\prime }\left( \tilde{q}^{2}\right) %
\right] }{\tilde{R}\left( \tilde{q}^{2}\right) +\Gamma _{\mathrm{int}%
}^{\left( 2\right) }\left[ \tilde{q};\tilde{M}_{1}\right] }+\mathcal{G}_{%
\mathrm{dil}}\left( \Gamma _{\mathrm{int}},\tilde{M},d_{\phi }^{\ast \left(
-\right) }\right) \,.  \label{eq:Morrisflow}
\end{equation}

\paragraph{The Wetterich procedure}

The cutoff function is assumed to display a global scale-de\-pen\-dent factor
noted $Z_{\Lambda }\left( e^{-t}\right) $, of the form:%
\begin{equation}
R\left( q^{2};\Lambda \right) =\Lambda ^{2n_{0}}Z_{\Lambda }\left(
e^{-t}\right) \tilde{R}\left( \tilde{q}^{2}\right) \,,  \label{eq:Rwett}
\end{equation}%
with [see eq.(\ref{eq:Z3}) in which $Z_{3}=1/Z_{\Lambda }$]\footnote{$%
Z_{\Lambda }\left( e^{-t}\right) $ is the inverse of the usual wave function
renormalization $Z_{3}\left( e^{-t}\right) $ which relates the
\textquotedblleft bare\textquotedblright\ field $\phi $ to the renormalized
field $\phi _{R}$ as $\phi =\sqrt{Z_{3}\left( e^{-t}\right) }\,\phi _{R}$.
Also, in \cite{4700},\ Wetterich's RG-time is the opposite of (\ref%
{eq:RGtime}), $n_{0}=1$, and $k$ and $\Lambda $ correspond respectively to
our $\Lambda $, $\mu _{0}$.\label{1surZ3}}:%
\begin{equation}
\eta \left( t\right) -2(1-n_{0})=-\Lambda \frac{\partial }{\partial \Lambda }%
Z_{\Lambda }\left( e^{-t}\right) =\frac{d}{dt}Z_{\Lambda }\left(
e^{-t}\right) \,,  \label{eq:etawett}
\end{equation}%
that, de facto introduces arbitrarily --within the cutoff function-- an
effect of a preliminar\footnote{%
Prior to the infinitesimal change of $\Lambda $ of interest to derive the RG
equation.} decimation over the finite range $[\Lambda ,\mu _{0}]$, see (\ref%
{eq:RGtime}). Then Wetterich defines a \textquotedblleft
dimensionless-renormalized\textquotedblright\ field $\tilde{M}_{1}$ as $%
M_{1}=\left[ Z_{\Lambda }\left( t\right) \right] ^{-1/2}\Lambda ^{d/2-n_{0}}%
\tilde{M}_{1}$ --corresponding to a field-renormalization over the finite
range $[\Lambda ,\mu _{0}]$ that compensates the previous effect (or perhaps
it is the reverse). It is easy to verify that, so doing, one obtains the
same equation as Morris (\ref{eq:Morrisflow}) with a significant difference,
however, that, this time, $\eta \left( t\right) $ is not necessarily fixed
to $\eta ^{\ast }$, but is allowed to flow (its flow is, in principle,
dictated by the constancy of the coefficient of the kinetic term of $\Gamma $
though it is not necessarily determined that way in \cite{4700}).

\paragraph{Back to the full action}

Because the explicit scale-dependence of the cutoff function now compensates
exactly the field-renormalisation, one may utilize the Legendre
transformation (\ref{eq0:TL1DimLess}, \ref{eq0:TL2DimLess}) to get, from (%
\ref{eq:Morrisflow}), the corresponding form of the RG-flow equation for $S_{%
\mathrm{int}}$ and then for\footnote{%
Having the flow for $S_{\mathrm{int}}$ one may use the relation (\ref%
{eq:SintDimLess}) back to $S$. This does not mean, however, that a Legendre
transformation links directly the flow equation satisfied by $S$ to that
satisfied by $\Gamma _{\mathrm{int}}$.} $S$. Obviously one does not recover (%
\ref{eq:PolchLin}) or (\ref{eq:PolchModif}) that way. Nevertheless one\
obtains a RG-flow equation for $S$ with a non-linear dependence on $\eta
\left( t\right) $ that may be directly compared with the historic Wilson
equation (or, rather, its version \textquotedblleft \textsl{extended to an
arbitrary cutoff function}\textquotedblright ). It appears that the
non-linear contribution proportional to $\eta \left( t\right) $ found that
way is opposite in sign compared to that of the historic first version (for
more detail, see \cite{7849}). As we shall see, this difference in sign is
due to the fact that the Morris-Wetterich procedure of introducing $\eta
\left( t\right) $, though perfectly acceptable, is artificial (forced) and
not directly associated to the realization of RG-step 1.

Is, this difference in sign, the reason why one has rather tried to adapt
the Legendre transformation to link the \textsl{linear} version of Ball et
al to the \textsl{nonlinear} version of Morris or Wetterich \cite%
{7289,7205,6750}? It is hard to answer that question but we may a priori
expect a complicated form of the Legendre transformation in that case.

\subsubsection{Elaborated Legendre transformation\label{Elab}}

Studies \cite{7289,7205,6750} have shown that one can define a Legendre
transformation that relates the \textquotedblleft
modified\textquotedblright\ Polchinski flow equation on the one hand to the
flow equation (\ref{eq:Morrisflow}) satisfied by $\Gamma _{\mathrm{int}}$
(even in the case of $\eta \left( t\right) $ not limited to be a constant 
\cite{7205}) on the other hand. Of course, because $\eta \left( t\right) $
is, notably, not introduced in the same way in the two flow equations, this
Legendre transformation is extremely complicated (but this is not the only
reason). It is obtained using a generalization of (\ref{eq0:TL1}, \ref%
{eq0:TL2}) written in terms of dimensionless quantities as follows:

\begin{eqnarray*}
S_{\mathrm{int}}\left[ \tilde{\phi}\right] &=&\Gamma _{\mathrm{int}}\left[ 
\tilde{M}\right] +\frac{1}{2}\tilde{M}\cdot \mathcal{R}\cdot \tilde{M}+\frac{%
1}{2}\tilde{\phi}\cdot \mathcal{Q}\cdot \tilde{\phi}-\tilde{\phi}\cdot 
\mathcal{L}\cdot \tilde{M}\,, \\
\mathcal{L}\cdot \tilde{M} &=&\mathcal{Q}\cdot \tilde{\phi}-\frac{\delta }{%
\delta \tilde{\phi}}S_{\mathrm{int}}\left[ \tilde{\phi}\right] \,,
\end{eqnarray*}%
in which the functions $\mathcal{R}\left( \tilde{q}^{2}\right) $, $\mathcal{Q%
}\left( \tilde{q}^{2}\right) $ and $\mathcal{L}\left( \tilde{q}^{2}\right) $
are adjusted so as to get (\ref{eq:Morrisflow}) from (\ref{eq:PolchModif}).
With a choice of cutoff function corresponding to (\ref{eq:defP}) and $%
n_{0}=1$, the solution reads \cite{7289}:%
\begin{eqnarray*}
\mathcal{Q}\left( x\right) &=&\frac{x}{K\left( x\right) }\left( \frac{1}{%
\sigma \left( x\right) }-1\right) \,, \\
\mathcal{L}\left( x\right) &=&\frac{x}{\sigma \left( x\right) }\,, \\
\mathcal{R}\left( x\right) &=&x\frac{K\left( x\right) }{\sigma \left(
x\right) }\,,
\end{eqnarray*}%
with:%
\begin{equation*}
\sigma \left( x\right) =K\left( x\right) x^{\eta ^{\ast
}/2}\int_{0}^{x}du\,u^{-\eta ^{\ast }/2}\,\frac{d}{du}\frac{1}{K\left(
u\right) }\,.
\end{equation*}

Of course, in the scale-de\-pen\-dent case $\eta \left( t\right) $, the
expressions are more involved \cite{7205}.

\section{Structural method\label{WilsonWay}}

\subsection{Reminder of the \textquotedblleft analytical
method\textquotedblright \label{ana0}}

Except for the historic first ERGE \cite{440}, the common way of obtaining
the RG-flow equations for $S_{\Lambda }$ is based on considering first $S_{%
\mathrm{int},\Lambda }$ and an explicit UV-cutoff function. Let us , call
\textquotedblleft analytical method\textquotedblright\ the procedure of
infinitesimally performing the three RG-steps first on $S_{\mathrm{int}%
,\Lambda }$. A normal analytic method may be sketched as follows (see part %
\ref{Sec:Sum}):

\begin{enumerate}
\item RG-step 1 is performed on $S_{\mathrm{int},\Lambda }$ through an
infinitesimal variation of the explicit UV-cutoff as introduced in (\ref%
{eq:Sint}),

\item the relation between $S_{\Lambda }$ and $S_{\mathrm{int},\Lambda }$ (%
\ref{eq:Sint}) is then considered to transfer the preceding result on $%
S_{\Lambda }$,

\item the rescaling RG-step 2 is (infinitesimally) performed both on $S$ and 
$S_{\mathrm{int}}$ to give the RG-flow equations for $S\left[ \tilde{\phi},t%
\right] $ and $S_{\mathrm{int}}\left[ \tilde{\phi},t\right] $ [this commutes
with the preceding step since the relation (\ref{eq:Sint}) becomes (\ref%
{eq:SintDimLess}) which has the same form and induces the equality (\ref%
{eq:Spt=SintPt}) $\dot{S}\left[ \tilde{\phi},t\right] =\dot{S}_{\mathrm{int}}%
\left[ \tilde{\phi},t\right] $],

\item the field-renormalization RG-step 3 is implemented linearly (i.e.
infinitesimally) within the RG-flow equation for either $S_{\mathrm{int}}%
\left[ \tilde{\phi},t\right] $ or $S\left[ \tilde{\phi},t\right] $,

\item the final relation that enables to deduce one RG-flow equation from
the other is altered compared to (\ref{eq:Spt=SintPt}), then one must modify
either the relation (\ref{eq:Spt=SintPt}) or one of the two RG-equations so
obtained (hence the so-called \textquotedblleft modified\textquotedblright\
Polchinski equation).
\end{enumerate}

A similar strategy could be sketched that characterizes a normal way of
treating the Legendre transformation linking the RG-flow equations for $S_{%
\mathrm{int},\Lambda }$ and $\Gamma _{\mathrm{int},\Lambda }$.

As indicated in sections \ref{WilERGE} and \ref{Implem}, Wilson on the one
hand and Morris-Wetterich on the other hand have not followed the
\textquotedblleft normal\textquotedblright\ way concerning the realization
of RG-step 3 since they have both used a non-infinitesimal field
renormalization. The Legendre transformation that links the Morris-Wetterich
RG-flow equation to the \textquotedblleft normal\textquotedblright\
Polchinski flow equation is very complicated.

Instead of adapting a Legendre transformation so as to obtain the
Morris-Wetterich RG-flow equation (\ref{eq:Morrisflow}) --as done in \cite%
{7289,7205,6750}, we aim at determining the flow equation satisfied by $%
\Gamma _{\mathrm{int}}$ that corresponds to a Wilsonian ERGE
\textquotedblleft \textsl{extended to an arbitrary cutoff function}%
\textquotedblright , while maintaining a simple form for the Legendre
transformation and for the link between $S$ and $S_{\mathrm{int}}$. Because
the method followed relies on structural properties of the ERGE, we call it
the structural method.

\subsection{Principle of the structural method\label{Prince}}

The basical idea is to implement RG-step 3 via an EMRO of the flow equation
for $S\left[ \tilde{\phi},t\right] $. Indeed, an EMRO is an
\textquotedblleft operator\textquotedblright\ responsible for the
infinitesimal change of field-normalization by a constant that moves the
action infinitesimally along a line of equivalent fixed points. Then, having
extended its expression out of the vicinity of fixed points, one may use it
to implement the field-renormalization step in the ERGE. This looks like the
Wilson procedure though the reasoning is inversed. In the historic first
version, the implementation of RG-step 3 is done via the operator that
actually infinitesimally changes the normalization of the field which, at a
fixed point and for the Wilson realization of the RG-step 1, coincides with
an EMRO (this is not always true, e.g. see \cite{7849} and appendix \ref%
{ObtainEMRO}). Of course we could continue to introduce RG-step 3 the same
way as Wilson did [i.e. using (\ref{eq:Ow1})]\ whatever the realization of
RG-step 1, but then a Legendre transformation as simple as that constructed
in this article might not be so easily obtained.

At first sight, it seems that it could be impossible to determine an EMRO
attached to a Wilsonian RG-flow equation which is incompletely known.
Fortunately, the operation of changing the normalization of the field by a
constant commutes with the RG-steps \cite{4421}. Then one may determine an
EMRO of a given RG-flow equation without having implemented RG-step 3 yet.
From this knowledge we may complete the RG-steps via the EMRO so determined,
being reassured that it will remain EMRO of the modified flow equation \cite%
{7849} (see also appendix \ref{ObtainEMRO}). From the general form (\ref%
{eq:Ogenw}) of a redundant operator, it is possible to determine a strategy
to find the expression of an EMRO attached to a Wilsoninan RG-flow equation 
\cite{6228} (see appendix \ref{ObtainEMRO}).

By extension of its definition, let us call also EMRO a solution of an
eigenvalue-RG-equation\footnote{%
Linearization of the RG flow in the vicinity of a fixed point.} with a zero
eigenvalue whatever the RG equation considered. Due to its universal
character, the whole spectrum of eigenvalues does not depend on the kind of
ERGE considered. For a scalar field and a given dimension $d$, the spectrum
is the same wether one considers the Wilson, the Polchinski or the Wetterich
ERGE. So, each kind of equation possesses an EMRO (in its extended meaning).
Then if one knows the expression of an EMRO for a given ERGE, one may deduce
the expression of the corresponding EMRO for any other ERGE provided that
relations between the various flow equations are given. The structural
method thus consists in finding the expressions of the various ERGE that
correspond to the Wilson-like ERGE within which RG-step 3 has been
implemented via an EMRO and for given simple (but justified) relations
between the various actions.

\subsection{The Wilson ERGE \textquotedblleft \textsl{extended to an
arbitrary cutoff function}\textquotedblright \label{Extended}}

The first step is to establish the expression of the Wilson ERGE --with the
field-renormalization implemented via an EMRO-- \textquotedblleft \textsl{%
extended to an arbitrary cutoff function}\textquotedblright\ \cite{5744,7849}%
.

\subsubsection{Structural derivation}

It is possible to deduce it, structurally, from the considerations of
sections \ref{WilERGE} and \ref{PolERGE}.

Considering a cutoff function $P\left( q^{2},\Lambda \right) $ as introduced
in (\ref{eq:Sint}), one may infer the following flow equation for $S\left[ 
\tilde{\phi},t\right] $, provided $n_{0}=0$:%
\begin{eqnarray}
\dot{S} &=&-\int_{\tilde{q}}\tilde{q}^{2}\tilde{P}^{\,\prime }\left( \tilde{q%
}^{2}\right) \left[ \frac{\delta ^{2}S}{\delta \tilde{\phi}_{\tilde{q}%
}\delta \tilde{\phi}_{-\tilde{q}}}-\frac{\delta S}{\delta \tilde{\phi}_{%
\tilde{q}}}\frac{\delta S}{\delta \tilde{\phi}_{-\tilde{q}}}+2\,\tilde{P}%
^{-1}\left( \tilde{q}^{2}\right) \,\tilde{\phi}_{\tilde{q}}\frac{\delta S}{%
\delta \tilde{\phi}_{\tilde{q}}}\right]  \notag \\
&&+\mathcal{G}_{\mathrm{dil}}\left( S,\tilde{\phi},d_{\phi }^{\left(
cw\right) }\right) +\varpi _{0}\left( t\right) \mathcal{O}_{P}\left( S,%
\tilde{\phi}\right) \,,  \label{eq:WilExtended}
\end{eqnarray}%
in which $d_{\phi }^{\left( cw\right) }$ is given by (\ref{eq:dcwil}) and $%
\mathcal{O}_{P}\left( S^{\ast },\tilde{\phi}\right) $ is an EMRO associated
to the flow equation linearized about a fixed point $S^{\ast }$. Written
under the general form (\ref{eq:Ogenw}), $\mathcal{O}_{P}\left( S,\tilde{\phi%
}\right) $ (considered out of the fixed point) corresponds to:%
\begin{equation}
\psi _{\tilde{q}}^{\left( P\right) }\left( \tilde{\phi},S\right) =\tilde{\phi%
}_{\tilde{q}}-\tilde{P}\left( \tilde{q}^{2}\right) \frac{\delta S}{\delta 
\tilde{\phi}_{-\tilde{q}}}\,,  \label{eq:PsiPS}
\end{equation}%
that is to say: 
\begin{equation}
\mathcal{O}_{P}\left( S,\tilde{\phi}\right) =\int_{\tilde{q}}\left[ \tilde{P}%
\left( \tilde{q}^{2}\right) \left( \frac{\delta ^{2}S}{\delta \tilde{\phi}_{%
\tilde{q}}\delta \tilde{\phi}_{-\tilde{q}}}-\frac{\delta S}{\delta \tilde{%
\phi}_{\tilde{q}}}\frac{\delta S}{\delta \tilde{\phi}_{-\tilde{q}}}\right) +%
\tilde{\phi}_{\tilde{q}}\frac{\delta S}{\delta \tilde{\phi}_{\tilde{q}}}-1%
\right] \,,  \label{eq:OP}
\end{equation}%
see \cite{7849} and appendix \ref{ObtainEMRO} for more detail.

Indeed, (\ref{eq:WilExtended}) is the flow equation for $S$ as obtained by
Ball et al \cite{3491}, using the analytical method sketched in section \ref%
{ana0}, but in which the field-renormalization step is implemented via $%
\mathcal{O}_{P}\left( S,\tilde{\phi}\right) $ instead of being implemented
linearly within $\mathcal{G}_{\mathrm{dil}}$ exclusively. Moreover, $%
\mathcal{O}_{P}\left( S,\tilde{\phi}\right) $ may be deduced from $\mathcal{O%
}_{w}\left( S,\tilde{\phi}\right) $ (\ref{eq:Ow1}) after the change\footnote{%
Though we could have expected that the EMRO be still the infinitesimal
realization of the operator $U_{\kappa }$, as given by (\ref{eq:U}), which
is responsible for a change of normalization of the field $\tilde{\phi}$.} $%
\tilde{\phi}\rightarrow \tilde{\phi}/\sqrt{\tilde{P}}$.

\begin{remark}
In order to preserve the quasi-local character of the action after the field-redefinition that allows to
recover the Wilson EMRO (\ref{eq:Ow1}), it is necessary that $n_{0}=0$. Consequently, in the following, if
not mentioned, we assume that $n_{0}=0$, i.e. $d_\phi^{\left(c\right)}\rightarrow d_\phi^{\left(cw\right)}$.
(Actually the EMRO for $n_{0}=1$ takes on a simple form also though it is a bit modified compared to (\ref{eq:OP})
in order to
preserve quasi-locality, see appendix \ref{ObtainEMRO}.) 
\end{remark}

If the above derivation of (\ref{eq:WilExtended}) is satisfactory, it is
complicated due to the necessity of determining first the expression of an
EMRO. Indeed we have been able to find (\ref{eq:OP}) because of the
preliminary knowledge of the original Wilson's EMRO (\ref{eq:Ow1}) which
itself was known to be constructed from the explicit transformation (\ref%
{eq:U}). However, because of the numerous possibilities of implementing
RG-step 1, and the freedom of redefining the field at will (the fourth step
of the RG-procedure), an EMRO is not always so directly related to (\ref%
{eq:Ow1}) (see appendix \ref{ObtainEMRO}). Moreover, the role of the cutoff
function, seen as a field redefinition, is partly hidden in this approach so
that it is instructive to look also at the \textquotedblleft
analytical\textquotedblright\ way of obtaining (\ref{eq:WilExtended}) based
on the consideration of an explicit UV-cutoff function. As already
encountered in section \ref{Implem}, the non-linear introduction of $\eta
\left( t\right) $ may be realized by assuming that the cutoff function
displays a factorized scale-dependence. In the following section we look at
this analytical way of obtaining (\ref{eq:WilExtended}) and
\textquotedblleft explain\textquotedblright\ how and why it produces a
contribution opposite in sign compared to the Morris-Wetterich way of
realizing RG-step 3. We will also observe that the recourse to an EMRO
implies, de facto, that the usual limit of infinite overall cutoff ($\Lambda
_{0}\rightarrow \infty $) be effected which finally justifies the simplicity
of the Legendre transformation (\ref{eq:001a}, \ref{eq:001b}) to be used in
the structural approach.

\subsubsection{Analytical derivation\label{ana}}

In \cite{5744} a derivation of (\ref{eq:WilExtended}) was given based on an
explicit ad-hoc scale-dependence within the cutoff function introduced by
the following change: 
\begin{eqnarray}
\tilde{P}\left( \tilde{q}^{2}\right) &\rightarrow &\zeta \left( t\right) 
\tilde{P}\left( \tilde{q}^{2}\right) \,,  \label{eq:ch1} \\
\zeta \left( t\right) &=&e^{-\left( 2-\eta ^{\ast }\right) t}\,.
\label{eq:ch2}
\end{eqnarray}

As already mentioned at the end of section \ref{Implem} and emphasized in 
\cite{7849}, the sign of the exponent in (\ref{eq:ch2}) is opposite to that
assumed by Morris and Wetterich within the cutoff function $R$ [see eqs (\ref%
{eq:RMorris}), (\ref{eq:Rwett}, \ref{eq:etawett}) with $n_{0}=0$ and $R\sim
1/\tilde{P}$]. An explanation of that latter point and a clear justification
of (\ref{eq:ch1}, \ref{eq:ch2}) was not given\footnote{%
In \cite{5744}, not only the origin of the change (\ref{eq:ch1}, \ref{eq:ch2}%
) was not strongly justified but also some misleading arguments have been
given. In particular a confusion has been made between $\zeta \left(
t\right) $ and the inverse of $Z_{3}\left( e^{-t}\right) $. The reason for
this confusion is \textquotedblleft explained\textquotedblright\ below.} in 
\cite{5744}. It is thus justified to present here a detailed re-derivation
and justification of (\ref{eq:WilExtended}) on the basis of a cutoff
function involving a factorized scale-dependence.

First, let us come back to the discussion of section \ref{WettERGE}
concerning the obtention of Polchinski's equation (\ref{eq:PolchBrute2})
from an IR cutoff function and its relation to the original procedure via (%
\ref{eq:Delta2P}, \ref{eq:Delta1P}). In original Polchinski's paper \cite%
{354}, the running cutoff $\Lambda $ is like an overall cutoff $\Lambda _{0}$%
, whereas the running cutoff $\Lambda $ of section \ref{WettERGE} is the
result of having integrated the degrees of freedom attached to the finite
momentum range $\left[ \Lambda ,\Lambda _{0}\right] $. Thus, for the
scale-de\-pen\-dent action $S_{\Lambda }\left[ \phi _{1}\right] $ to be like the
Polchinski action, two steps have still to be carried out:

\begin{enumerate}
\item the rescaling $\Lambda \rightarrow \Lambda _{0}$

\item the field-renormalization $\phi _{1}\rightarrow \sqrt{Z_{3}\left( 
\frac{\mu }{\Lambda }\right) }\phi _{R}$ with $\mu <\Lambda $.
\end{enumerate}

Before going any further, let us draw attention to the following subtleties:

\begin{enumerate}
\item[a.] in the system of units set by the overall cutoff $\Lambda _{0}$, $%
\mu $ represents the momentum-scale value reached by the running scale $%
\Lambda $ \textsl{after} decimation and \textsl{before} rescaling. After
rescaling, the running scale $\Lambda $ takes on the initial value $\Lambda
_{0}$ whereas the overall cutoff $\Lambda _{0}$ is increased at the same
rate. Obviously, the ratio $\mu /\Lambda $ has the same magnitude as the
ratio $\Lambda /\Lambda _{0}$, but the field-renormalization function $Z_{3}$
depends on the ratio $\mu /\Lambda $ (corresponding to the range over which
the decimation is effected) and not on the ratio $\Lambda /\Lambda _{0}$
(contrary to Wetterich's procedure of field-renormalization\footnote{%
Although, in Wetterich's procedure, $\Lambda _{0}$ has already been sent to
infinity and is implicitly replaced by and arbitrary momentum-scale $\mu
_{0} $, the argument is not altered.}, see section \ref{Implem}). Thus, the
dependence on the running scale $\Lambda $ within $Z_{3}\left( \frac{\mu }{%
\Lambda }\right) $ occurs rightly, but unusually [e.g., in comparison to (%
\ref{eq:etawett})], in the denominator of the scale ratio so that, according
to (\ref{eq:Z3}), $\Lambda \frac{d}{d\Lambda }Z_{3}\left( \frac{\mu }{%
\Lambda }\right) =-\ell \frac{d}{d\ell }Z_{3}\left( \ell \right) =\frac{d}{dt%
}Z_{3}\left( e^{-t}\right) =2\varpi _{0}\left( t\right) Z_{3}$.

\item[b.] the usual limit of infinite initial cutoff $\Lambda
_{0}\rightarrow \infty $ at fixed $\Lambda $ induces $\mu \rightarrow 0$.
That is to say, taking that limit would be equivalent to assuming that, in
some sense, the RG program would have already been entirely (pre)-effected
over the whole momentum range $\left[ 0,\infty \right[ $ whereas $\Lambda $
would remain finite and non zero! We will see that this apparently strange
procedure concerns only the field-renormalization and is obliged in order to
\textquotedblleft analytically explain\textquotedblright\ the implementation
of RG-step 3 via an EMRO.
\end{enumerate}

For now we still assume that $\Lambda _{0}$ is finite. Then, the completion
of the field-renormalization step over the finite range $\left[ \mu ,\Lambda %
\right] $ implies that (\ref{eq:Delta1P}) becomes:%
\begin{equation}
P\left( q^{2},\Lambda \right) =\left[ Z_{3}\left( \frac{\mu }{\Lambda }%
\right) \right] ^{-1}\Delta _{1}\left( q^{2},\Lambda \right) \,.
\label{eq:PDelta1}
\end{equation}

If, following Polchinski, one considers that $P\left( q^{2},\Lambda \right) $
corresponds to the true origin of the RG-time, then $\Lambda =\Lambda _{0}$
and thus $\frac{\mu }{\Lambda }=1$ that is to say $Z_{3}=1$, consequently (%
\ref{eq:PDelta1}) coincides with (\ref{eq:Delta1P}). Notice that\
considering the limit $\Lambda _{0}\rightarrow \infty $ has no meaning in
the circumstances. That is the genuine Polchinski starting point.

Instead, but equivalently, one can imagine that it exists an overall cutoff $%
\Lambda _{0}>$ $\Lambda $ consequence of the implementation of a complete RG
procedure (three steps) over the finite range $\left[ \Lambda ,\Lambda _{0}%
\right] $ and preliminary to the infinitesimal reduction of $\Lambda $. One
may still consider that, nevertheless, $P\left( q^{2},\Lambda \right) $ is
unchanged compared to the original Polchinski version (after the RG
procedure the system is unchanged except for a reduction of the momentum
range). In that case the relation (\ref{eq:PDelta1}) would differ from (\ref%
{eq:Delta1P}) --because of the decimation that would have been actually
effected over the range $\left[ \mu ,\Lambda \right] $ with $\mu <\Lambda $.
Indeed, to keep $P\left( q^{2},\Lambda \right) $ unchanged after the field
renormalization over the range $\left[ \mu ,\Lambda \right] $, $\Delta
_{1}\left( q^{2},\Lambda \right) $ must have acquired --prior to field
renormalization-- a factorized scale-dependence\footnote{%
Which may be seen as consequence of the decimation --as it stands for any
coefficient of the scale-de\-pen\-dent action $S\left[ \tilde{\phi},t\right] $
prior to field-renormalization.} given by the inversion of (\ref{eq:PDelta1}%
):%
\begin{equation}
\Delta _{1}\left( q^{2},\Lambda \right) =\left[ Z_{3}\left( \frac{\mu }{%
\Lambda }\right) \right] P\left( q^{2},\Lambda \right) \,.
\label{eq:PDelta1bis}
\end{equation}

Clearly, in terms of $\phi _{R}$, Polchinski's procedure would remain
unchanged and his equation would not be modified. Nevertheless, this time, a
totally ineffective finite overall cutoff $\Lambda _{0}$ would (implicitly)
exist --together with a possible implicit IR-momentum-scale $\mu $, see
below. However, these latter two scales do not alter the RG-flow because
only the local (infinitesimal) variation of $\Lambda $ is of interest.

From this, we may foresee a new point of view. Let us suppose that instead
of effecting the three RG-steps over the finite range $\left[ \Lambda
,\Lambda _{0}\right] $ \textsl{before} considering the construction of the
Polchinski ERGE (via an infinitesimal variation of $\Lambda $), we assume
that only RG-step 1 and 2 have been effected. Thus, in order to complete the
three RG-steps within the ERGE, the field-renormalization will have still to
be effected globally over the finite range $\left[ \mu ,\Lambda \right] $
instead of being implemented only infinitesimally. Then, to start with the
derivation of the ERGE (via an infinitesimal variation of $\Lambda $), let
us suppose that the field of reference is the \textquotedblleft
unrenormalized\textquotedblright\ field $\phi _{1}$ (the \textquotedblleft
old field\textquotedblright\ would say Bell and Wilson \cite{4421}).
Consequently, Eq. (\ref{eq:PDelta1bis}) defines $\Delta _{1}\left(
q^{2},\Lambda \right) $ in terms of the regular UV cutoff function $P\left(
q^{2},\Lambda \right) $. In that case, (\ref{eq:Delta2}) reads:

\begin{equation}
\Delta _{2}\left( q^{2};\Lambda ,\Lambda _{0}\right) =\Delta _{0}\left(
q^{2},\Lambda _{0}\right) -\left[ Z_{3}\left( \frac{\mu }{\Lambda }\right) %
\right] P\left( q^{2},\Lambda \right) \,,  \label{eq:Delta2bis}
\end{equation}%
so that the derivative w.r.t. $\Lambda $ in the Polchinski equation (\ref%
{eq:PolchBrute2}) generates a contribution proportional to $\left[ \Lambda 
\frac{d}{d\Lambda }Z_{3}\left( \frac{\mu }{\Lambda }\right) \right] =2\varpi
_{0}\left( t\right) Z_{3}\left( \frac{\mu }{\Lambda }\right) $ which has a
sign opposite to what would be obtained with the Morris-Wetterich procedures
described in section \ref{Implem}. We emphasize again that this change of
sign is not due to Wetterich's choice of $Z_{\Lambda }=1/Z_{3}$ but rather
to the fact that $\Lambda $ occurs \textsl{unusually, but correctly}, in the
denominator of the scale ratio within $Z_{3}$.

The procedure for obtaining the ERGE is unchanged relatively to the first
two steps: an infinitesimal change $\Lambda \rightarrow \Lambda -d\Lambda $
with its associated infinitesimal rescaling (with a classical $\mathcal{G}_{%
\mathrm{dil}}$); however the third step (field-renormalization) has to be
realized over the full finite range $\left[ \mu ,\Lambda \right] $ instead
of being implemented infinitesimally and this induces a subtlety in the
procedure. Actually the factor $\left[ Z_{3}\left( \frac{\mu }{\Lambda }%
\right) \right] $ in (\ref{eq:Delta2bis}) has anticipated the \textsl{%
consequence} of the infinitesimal realization of the RG-step 1 over the
range $\left[ \Lambda -d\Lambda ,\Lambda \right] \subset $ $\left[ \mu
,\Lambda \right] $ which is under present consideration. Hence, this part of 
$\left[ Z_{3}\left( \frac{\mu }{\Lambda }\right) \right] $ must be taken
away from the derivative w.r.t. $\Lambda $. So, to avoid an overlapping
between the $\Lambda $-dependencies in $Z_{3}$ and $P$ (i.e., to avoid a
possible double counting), we must effectuate a \textquotedblleft
pre-renormalization\textquotedblright\ of the field ($\phi \rightarrow \sqrt{%
Z_{3}\left( \frac{\Lambda -d\Lambda }{\Lambda }\right) }\phi $) over the
infinitesimal range\footnote{%
As we did it above, over the whole range $\left[ \mu ,\Lambda \right] $, to
recover Polchinski's procedure.} since then the infinitesimal RG-step 1 will
regenerate the full factor $Z_{3}\left( \frac{\mu }{\Lambda }\right) $ in
front of $P$. After that, the final field-renormalization can be truly
performed over the full range $\left[ \mu ,\Lambda \right] $. Finally, after
completing the field renormalization, but not yet the infinitesimal
rescaling, we get: 
\begin{eqnarray}
\left. \Lambda \frac{\partial }{\partial \Lambda }S_{\mathrm{int}}\left[
\phi \right] _{\Lambda }\right\vert _{\phi } &=&-\frac{1}{2}\int_{q}\left\{ %
\left[ 2\varpi _{0}\left( t\right) P\left( q^{2};\Lambda \right) +\left.
\Lambda \frac{\partial P\left( q^{2};\Lambda \right) }{\partial \Lambda }%
\right\vert _{q}\right] \right.  \notag \\
&&\left. \times \left[ \frac{\delta ^{2}S_{\mathrm{int}}\left[ \phi \right]
_{\Lambda }}{\delta \phi _{q}\delta \phi _{-q}}-\frac{\delta S_{\mathrm{int}}%
\left[ \phi \right] _{\Lambda }}{\delta \phi _{q}}\frac{\delta S_{\mathrm{int%
}}\left[ \phi \right] _{\Lambda }}{\delta \phi _{-q}}\right] \right\}  \notag
\\
&&+\varpi _{0}\left( t\right) \,\phi \cdot \frac{\delta }{\delta \phi }S_{%
\mathrm{int}}\left[ \phi \right] _{\Lambda }\,,  \label{eq:PolchBrute3}
\end{eqnarray}%
in which $\phi $ stands for $\phi _{R}$ and the last term is due to the
required \textquotedblleft pre-renormalization\textquotedblright\ with $%
\varpi _{0}\left( t\right) $ given by (\ref{eq:Z3}).

After rescaling we obtain:%
\begin{eqnarray}
\dot{S}_{\mathrm{int}}\left[ \tilde{\phi},t\right] &=&\int_{\tilde{q}}\left(
\varpi _{0}\left( t\right) \tilde{P}\left( \tilde{q}^{2}\right) -\tilde{q}%
^{2}\tilde{P}^{\,\prime }\left( \tilde{q}^{2}\right) \right) \left[ \frac{%
\delta ^{2}S_{\mathrm{int}}}{\delta \tilde{\phi}_{\tilde{q}}\delta \tilde{%
\phi}_{-\tilde{q}}}-\frac{\delta S_{\mathrm{int}}}{\delta \tilde{\phi}_{%
\tilde{q}}}\frac{\delta S_{\mathrm{int}}}{\delta \tilde{\phi}_{-\tilde{q}}}%
\right]  \notag \\
&&+\mathcal{G}_{\mathrm{dil}}\left( S_{\mathrm{int}},\tilde{\phi},d_{\phi
}^{\left( +\right) }\left( t\right) \right) \,,  \label{eq:PolchStruct}
\end{eqnarray}%
in which $d_{\phi }^{\left( +\right) }\left( t\right) $, defined in (\ref%
{eq:dphi+-}), differs from the usual $d_{\phi }^{\left( -\right) }\left(
t\right) $, but is quite warranted.

Then it is easy to verify that the Wilsonian RG-flow (\ref{eq:WilExtended})
--under its exact writing-- may be obtained either from (\ref{eq:PolchBrute3}%
) and the usual relation (\ref{eq:Sint}) --with the rescaling step finally
completed on $S$-- or directly from (\ref{eq:PolchStruct}) and the
dimensionless form (\ref{eq:SintDimLess}) --with the condition (\ref%
{eq:Spt=SintPt}).

\subsubsection{Additional important comment\label{MuVers0}}

Though neither $\mu $ nor $\Lambda _{0}$ appear explicitly in the RG-flow
equations for $S$ and $S_{\mathrm{int}}$ obtained in the preceding section,
the analytical method followed relies upon an explicit reference to them
(via $\mu $ occuring within the cutoff function). Contrary to $\Lambda _{0}$
that obviously has no effect on the usual Polchinski RG-flow equation, the
reference to $\mu $ is not completely harmless. Remind that $\mu \neq 0$ is
a consequence of implementing the field-renormalization corresponding to the
decimation over the range $\left[ \Lambda ,\Lambda _{0}\right] $ (i.e. those
fields having momentum components in the finite range $\left[ \mu ,\Lambda %
\right] $ with $\mu <\Lambda $). Then, because the RG-flow corresponds to a
decreasing of $\Lambda $, it is almost obvious that $\mu $ is formally an
unacceptable\ lower limit to the running scale $\Lambda $. This is because
for $\Lambda <\mu $ the field-renormalization is no longer accounted for
within $Z_{3}(\mu /\Lambda )$ and, thus, should be completed the usual
linear way (what has not been expressed in the ERGE). Though $\mu $ does not
appear in our writing of the RG-flow equation, it should\footnote{%
For the same reason we should mention that the ERGE is only valid for $%
\Lambda <\Lambda _{0}$, what is never done with a Polchinski like equation
in which $\Lambda _{0}$ never appears explicitly.}. Thus, strictly speaking,
the analytical method reveals that the RG-flow equation in which RG-step 3
is implemented via an EMRO either is only valid\footnote{%
This limitation does not occur in the Morris-Wetterich way of implementing
RG-step 3 because it is based on the introduction of the ad-hoc factor $%
Z_{3}\left( \Lambda /\mu _{0}\right) $ in front of the cutoff function with $%
\mu _{0}>\Lambda $.} for $\Lambda >\mu $ or implies the limit $\mu
\rightarrow 0$ be effected. Now $\mu $ plays the role of an overall
IR-cutoff that is the counter-part of the UV-cutoff $\Lambda _{0}$. It is
almost obvious that, to make a complete contact with the Wilson procedure of
implementing RG-step 3 using an EMRO at any scale, the limit $\mu
\rightarrow 0$ must be explicitly achieved (together with the concomitant
limit $\Lambda _{0}\rightarrow \infty $). This point, that may be connected
to remark \ref{Bell}, will be made even clearer with the consideration of
the RG-flow equation for the effective action $\Gamma $, see sections \ref%
{EMROWett} and \ref{Limite}.

\subsection{Structural constructions\label{Struc}}

Because the simple Legendre transformation (\ref{eq:001a},\ref{eq:001b}) is
not yet justify, we still need to use partly the analytical version
requiring to consider first $S_{\mathrm{int}}$ and $\Gamma _{\mathrm{int}}$
instead of considering exclusively to the full actions $S$ and $\Gamma $.
Thus, having justified both structurally and analytically\footnote{%
At least for $\mu <\Lambda $, see section \ref{MuVers0}.} the form of (\ref%
{eq:WilExtended}) we may lean on it to rederive structurally the
Polchinski-like ERGE (\ref{eq:PolchStruct}) and, via the usual Legendre
transform, the corresponding Wetterich-like ERGE.

\subsubsection{Polchinski-like ERGE\label{EMROPol}}

After implementation of the three RG-steps, and owing to the factorized
scale-dependence of the cutoff function (\ref{eq:PDelta1bis}) that entirely
absorbs the global field-renormalization step over the range $\left[ \mu
,\Lambda \right] $, we may assume the relations (\ref{eq:SintDimLess}, \ref%
{eq:Spt=SintPt}) between $S\left[ \tilde{\phi},t\right] $ and $S_{\mathrm{int%
}}\left[ \tilde{\phi},t\right] $. Then, considering the expression (\ref%
{eq:OP}) for the EMRO of the extended Wilson ERGE (\ref{eq:WilExtended}), we
may easily find its expression in terms of $S_{\mathrm{int}}$ using (\ref%
{eq:SintDimLess}). The resulting expression should be the expression of the
EMRO associated to the Polchinski ERGE, it comes \cite{7849}:

\begin{equation}
\mathcal{O}_{P}\left( S_{\mathrm{int}},\tilde{\phi}\right) =\int_{\tilde{q}}%
\left[ \tilde{P}\left( \tilde{q}^{2}\right) \left( \frac{\delta ^{2}S_{%
\mathrm{int}}}{\delta \tilde{\phi}_{\tilde{q}}\delta \tilde{\phi}_{-\tilde{q}%
}}-\frac{\delta S_{\mathrm{int}}}{\delta \tilde{\phi}_{\tilde{q}}}\frac{%
\delta S_{\mathrm{int}}}{\delta \tilde{\phi}_{-\tilde{q}}}\right) -\tilde{%
\phi}_{\tilde{q}}\frac{\delta S_{\mathrm{int}}}{\delta \tilde{\phi}_{\tilde{q%
}}}\right] \,,  \label{eq:OPSint}
\end{equation}

Finally, from the general property:%
\begin{equation}
\mathcal{G}_{\mathrm{dil}}\left( \frac{1}{2}\tilde{\phi}\cdot \tilde{P}%
^{-1}\cdot \tilde{\phi},\tilde{\phi},d_{\phi }^{\left( c\right) }\right)
=\int_{\tilde{q}}\tilde{\phi}_{\tilde{q}}\left[ n_{0}\tilde{P}^{-1}-\tilde{q}%
^{2}\left( \tilde{P}^{-1}\right) ^{\prime }\right] \tilde{\phi}_{-\tilde{q}%
} \,,  \label{eq:Gdilquadri}
\end{equation}%
in which we actually set $n_{0}=0$, it is easy to verify that the ERGE
satisfied by $S_{\mathrm{int}}$ that accounts for the three RG-steps
structurally deduced from (\ref{eq:WilExtended}) reads \cite{7849}: 
\begin{eqnarray}
\dot{S}_{\mathrm{int}} &=&-\int_{\tilde{q}}\tilde{q}^{2}\tilde{P}^{\,\prime
}\left( \tilde{q}^{2}\right) \left[ \frac{\delta ^{2}S_{\mathrm{int}}}{%
\delta \tilde{\phi}_{\tilde{q}}\delta \tilde{\phi}_{-\tilde{q}}}-\frac{%
\delta S_{\mathrm{int}}}{\delta \tilde{\phi}_{\tilde{q}}}\frac{\delta S_{%
\mathrm{int}}}{\delta \tilde{\phi}_{-\tilde{q}}}\right]  \notag \\
&&+\mathcal{G}_{\mathrm{dil}}\left( S_{\mathrm{int}},\tilde{\phi},d_{\phi
}^{\left( cw\right) }\right) +\varpi _{0}\left( t\right) \,\mathcal{O}%
_{P}\left( S_{\mathrm{int}},\tilde{\phi}\right) \,.  \label{eq:SintPoint}
\end{eqnarray}

If one splits the contribution proportionnal to $\mathcal{O}_{P}\left( S_{%
\mathrm{int}},\tilde{\phi}\right) $ in two parts, we then recover (\ref%
{eq:PolchStruct}) as it must.

It is also interesting to express the general form of a redundant operator (%
\ref{eq:Ogenw}) in terms of $S_{\mathrm{int}}$, it comes:%
\begin{equation}
\mathcal{O}\left( S_{\mathrm{int}},\tilde{\phi}\right) =\int_{\tilde{q}}%
\left[ \bar{\psi}_{\tilde{q}}\left( \tilde{\phi},S_{\mathrm{int}}\right)
\left( \tilde{P}^{-1}\left( \tilde{q}^{2}\right) \tilde{\phi}_{-\tilde{q}}+%
\frac{\delta S_{\mathrm{int}}}{\delta \tilde{\phi}_{\tilde{q}}}\right) -%
\frac{\delta \bar{\psi}_{\tilde{q}}\left( \tilde{\phi},S_{\mathrm{int}%
}\right) }{\delta \tilde{\phi}_{\tilde{q}}}\right] \,.  \label{eq:OSint}
\end{equation}

Thus (\ref{eq:OPSint}) corresponds to: 
\begin{equation}
\bar{\psi}_{\tilde{q}}^{\left( P\right) }\left( \tilde{\phi},S_{\mathrm{int}%
}\right) =-\tilde{P}\left( \tilde{q}^{2}\right) \frac{\delta S_{\mathrm{int}}%
}{\delta \tilde{\phi}_{-\tilde{q}}}\,.  \label{eq:PsiPSint}
\end{equation}

\subsubsection{Wetterich-like ERGE\label{EMROWett}}

Similarly to the preceding section, we may deduce from (\ref{eq:OPSint}) the
expression of the EMRO in terms of $\Gamma _{\mathrm{int}}$. To this end we
use the Legendre transformation (\ref{eq0:TL1}, \ref{eq0:TL2}) rewritten in
terms of renormalized dimensionless quantities (i.e. after rescaling and
field-renormalization) that reads:%
\begin{eqnarray}
S_{\mathrm{int}}\left[ \tilde{\phi},t\right] &=&\frac{1}{2}\left( \tilde{M}-%
\tilde{\phi}\right) \cdot \bar{\Delta}_{2}^{-1}\cdot \left( \tilde{M}-\tilde{%
\phi}\right) +\Gamma _{\mathrm{int}}\left[ \tilde{M},t\right] \,,
\label{eq:TL1sansDim} \\
\tilde{M} &=&\tilde{\phi}-\bar{\Delta}_{2}\cdot \frac{\delta }{\delta \tilde{%
\phi}}S_{\mathrm{int}}\left[ \tilde{\phi},t\right] \,,  \label{eq:TL2sansDim}
\\
\dot{S}_{\mathrm{int}}\left[ \tilde{\phi},t\right] &=&\dot{\Gamma}_{\mathrm{%
int}}\left[ \tilde{M},t\right] \,,  \label{eq:TL3sansDim}
\end{eqnarray}%
in which $\tilde{M}$ is a \textquotedblleft dimensionless
renormalized\textquotedblright\ field defined similarly to $\tilde{\phi}$
and 
\begin{equation*}
\bar{\Delta}_{2}\left( \tilde{q}^{2},t\right) =\left[ Z_{3}\left( \frac{\mu 
}{\Lambda }\right) \right] ^{-1}\tilde{\Delta}_{2}\left( \tilde{q}^{2},t,%
\frac{\Lambda _{0}}{\mu _{0}}\right) \,,
\end{equation*}%
is the dimensionless IR-cutoff function after field-renormalization. Using (%
\ref{eq:Delta2}) and (\ref{eq:Delta2bis}), it reads also:%
\begin{equation}
\bar{\Delta}_{2}\left( \tilde{q}^{2},t\right) =\left[ Z_{3}\left( \frac{\mu 
}{\Lambda }\right) \right] ^{-1}\tilde{\Delta}_{0}\left( \tilde{q}^{2},\frac{%
\Lambda }{\Lambda _{0}}\right) -\tilde{P}\left( \tilde{q}^{2}\right) \,.
\label{eq:Delta2bar}
\end{equation}

The expression of the EMRO (\ref{eq:OPSint}) thus leads to:%
\begin{eqnarray}
\mathcal{O}_{P}\left( \Gamma _{\mathrm{int}},\tilde{M}\right) =\int_{\tilde{q%
}} &&\left\{ \left[ \tilde{P}\left( \tilde{q}^{2}\right) \frac{\Gamma _{%
\mathrm{int}}^{\left( 2\right) }\left[ \tilde{q};\tilde{M}\right] }{1+\bar{%
\Delta}_{2}\Gamma _{\mathrm{int}}^{\left( 2\right) }\left[ \tilde{q};\tilde{M%
}\right] }\right. \right.   \notag \\
&&\left. \left. -\left( \tilde{P}\left( \tilde{q}^{2}\right) +\bar{\Delta}%
_{2}\left( \tilde{q}^{2},t\right) \right) \frac{\delta \Gamma _{\mathrm{int}}%
}{\delta \tilde{M}_{\tilde{q}}}\frac{\delta \Gamma _{\mathrm{int}}}{\delta 
\tilde{M}_{-\tilde{q}}}\right] -\tilde{M}_{\tilde{q}}\frac{\delta \Gamma _{%
\mathrm{int}}}{\delta \tilde{M}_{\tilde{q}}}\right\}   \label{eq:EMROWett1}
\end{eqnarray}%
and the flow equation for $\Gamma _{\mathrm{int}}$, deduced from (\ref%
{eq:PolchStruct}) or (\ref{eq:SintPoint}), reads: 
\begin{eqnarray}
\dot{\Gamma}_{\mathrm{int}}\left[ \tilde{M}\right] &=&-\int_{q}q^{2}\tilde{P}%
^{\,\prime }\left( q^{2}\right) \left\{ \frac{\Gamma _{\mathrm{int}}^{\left(
2\right) }\left[ \tilde{q};\tilde{M}\right] }{1+\bar{\Delta}_{2}\Gamma _{%
\mathrm{int}}^{\left( 2\right) }\left[ \tilde{q};\tilde{M}\right] }\right\} 
\notag \\
&&+\int_{q}\left\{ q^{2}\left[ \bar{\Delta}_{2}^{\prime }\left( \tilde{q}%
^{2},t\right) +\tilde{P}^{\,\prime }\left( q^{2}\right) \right] \frac{\delta
\Gamma _{\mathrm{int}}}{\delta \tilde{M}_{-q}}\,\frac{\delta \Gamma _{%
\mathrm{int}}}{\delta \tilde{M}_{q}}\right\}  \notag \\
&&+\mathcal{G}_{\text{\textrm{dil}}}\left( \Gamma _{\mathrm{int}},\tilde{M}%
,d_{\phi }^{\left( cw\right) }\right) +\varpi _{0}\left( t\right) \mathcal{O}%
_{P}\left( \Gamma _{\mathrm{int}},\tilde{M}\right) \,,
\label{eq:RGflowGammaInt0}
\end{eqnarray}%
in which $\bar{\Delta}_{2}^{\prime }\left( \tilde{q}^{2},t\right) =\left.
\partial \bar{\Delta}_{2}\left( \tilde{q}^{2},t\right) /\partial \tilde{q}%
^{2}\right\vert _{t}$. To get this results we have used the property that,
for $n_{0}=0$:%
\begin{equation*}
\mathcal{G}_{\text{\textrm{dil}}}\left( S_{\mathrm{int}},\tilde{\phi}%
,d_{\phi }^{\left( cw\right) }\right) =\mathcal{G}_{\text{\textrm{dil}}%
}\left( \Gamma _{\mathrm{int}},\tilde{M},d_{\phi }^{\left( cw\right)
}\right) +\int_{\tilde{q}}\left\{ \tilde{q}^{2}\bar{\Delta}_{2}^{\prime
}\left( \tilde{q}^{2},t\right) \frac{\delta \Gamma }{\delta \tilde{M}_{-%
\tilde{q}}}\,\frac{\delta \Gamma }{\delta \tilde{M}_{\tilde{q}}}\right\} \,.
\end{equation*}

Similarly to the preceding section, it is interesting to express, for given $%
\tilde{P}$ and $\bar{\Delta}_{2}$ the general form of a redundant operator (%
\ref{eq:Ogenw}) in terms of $\Gamma _{\mathrm{int}}$, it comes:%
\begin{eqnarray*}
\mathcal{O}\left( \Gamma _{\mathrm{int}},\tilde{M}\right) =\int_{\tilde{q}}
&&\left\{ \bar{\Psi}_{\tilde{q}}\left( \tilde{M},\Gamma _{\mathrm{int}%
}\right) \left[ \tilde{P}^{-1}\left( \tilde{q}^{2}\right) \left( \tilde{M}_{-%
\tilde{q}}+\bar{\Delta}_{2}\frac{\delta \Gamma _{\mathrm{int}}}{\delta 
\tilde{M}_{\tilde{q}}}\right) +\frac{\delta \Gamma _{\mathrm{int}}}{\delta 
\tilde{M}_{\tilde{q}}}\right] \right. \\
&&-\left. \frac{1}{1+\bar{\Delta}_{2}\Gamma _{\mathrm{int}}^{\left( 2\right)
}\left[ \tilde{q};\tilde{M}\right] }\frac{\delta \bar{\Psi}_{\tilde{q}%
}\left( \tilde{M},\Gamma _{\mathrm{int}}\right) }{\delta \tilde{M}_{\tilde{q}%
}}\right\} \,.
\end{eqnarray*}

Thus (\ref{eq:EMROWett1}) corresponds to: 
\begin{equation*}
\bar{\Psi}_{\tilde{q}}^{\left( P\right) }\left( \tilde{M},\Gamma _{\mathrm{%
int}}\right) =-\tilde{P}\left( \tilde{q}^{2}\right) \frac{\delta \Gamma _{%
\mathrm{int}}}{\delta \tilde{M}_{-\tilde{q}}}\,.
\end{equation*}

One observes that the expression of the ERGE is complicated due to the
explicit dependence on $t$ carried by $\bar{\Delta}_{2}$. Usually one get
rid of such an explicit dependence after completing the limit $\Lambda
_{0}\rightarrow \infty $ at fixed $\Lambda $.

\subsubsection{Limit of infinite overall cutoff\label{Limite}}

In the limit $\Lambda _{0}\rightarrow \infty $ the overall-cutoff function $%
\Delta _{0}$ is commonly assumed to approach a regular form (according to
the classical dimension (\ref{eq:dc}) of $\phi $):%
\begin{equation*}
\lim_{\Lambda _{0}\rightarrow \infty }\tilde{\Delta}_{0}\left( \tilde{q}^{2},%
\frac{\Lambda }{\Lambda _{0}}\right) =\frac{a}{\left( \tilde{q}^{2}\right)
^{n_{0}}}\,,
\end{equation*}%
so that, keeping $\Lambda $ and $\mu $ fixed, we have: 
\begin{equation}
\lim_{\Lambda _{0}\rightarrow \infty }\bar{\Delta}_{2}\left( \tilde{q}%
^{2},t\right) =\left[ Z_{3}\left( \frac{\mu }{\Lambda }\right) \right] ^{-1}%
\frac{a}{\left( \tilde{q}^{2}\right) ^{n_{0}}}-\tilde{P}\left( \tilde{q}%
^{2}\right) \,.  \label{eq:DeltaInf0}
\end{equation}

This result shows again a $t$ dependence within $\bar{\Delta}_{2}$ that
illustrates perfectly our purposes in section \ref{MuVers0} that $\mu $ is
an unacceptable lower limit to the running scale $\Lambda $. This is why we
are reassured by the fact that the commonly autorized limit $\Lambda
_{0}\rightarrow \infty $ induces in fact also the limit $\mu \rightarrow 0$
(see section \ref{ana}). Actually, the reasoning should be the reverse
because to justify the recourse to an EMRO we should perform the limit $\mu
\rightarrow 0$ which, then, would induce the limit $\Lambda _{0}\rightarrow
\infty $. Indeed, the nature of\ the origin of $\mu $ is different from that
of $\Lambda _{0}$: the scale $\mu $ appears for technical reasons and is
linked to the particular way we have implemented RG-step 3 whereas $\Lambda
_{0}$ may have a physical origin that sometimes we would like to keep (not
for field theoretical purposes however). It is interesting that the
necessity of considering $\mu \rightarrow 0$, which seems to be inherent to
the structural method, implies that $\Lambda _{0}$ must go to infinity. But
we are more accustomed to the process of sending $\Lambda _{0}$ to infinity
alone. So, to determine the value reached by $Z_{3}\left( \frac{\mu }{%
\Lambda }\right) $ when $\mu \rightarrow 0$, let us recall the conditions
under which we are commonly autorized to send $\Lambda _{0}$ to infinity.

The main interest of the RG in field theory is that, on decreasing $\Lambda $%
, the RG-flow may approach eventual fixed points (or infra-red stable
submanifolds) in the evolution-space of the actions. Such approaches in
their final evolution are so slow and the RG-time value is so large that the
initial $\Lambda _{0}$ is very large compared to $\Lambda $. Moreover the
precise form of $\Delta _{0}$ has practically no effect anymore on the
RG-flows in the vicinity of IR-stable manifolds, only the local topology of
the RG trajectories in the space of actions is important. Then, provided
that the RG-equation remains well defined, we may safely send $\Lambda _{0}$
to infinity. Said in other words: only the close vicinity of IR-stable
submanifilds are of interest in field theory so that one may get rid of both 
$\mu $ and $\Lambda _{0}$. The knowledge of the resulting RG-flow equation
is sufficient to determine the existence and the nature of the fixed points
of interest for field theory. Consequently to determine the behavior of $%
Z_{3}\left( \frac{\mu }{\Lambda }\right) $ in the limit $\mu \rightarrow 0$
at fixed $\Lambda $, of interest for field theory, it is necessary to assume
the vicinity of a fixed point.

According to (\ref{eq:Z3}, \ref{eq:cdet}), near a fixed point and for $%
n_{0}=0$, we expect $Z_{3}\left( \ell \right) $ to behave as:%
\begin{equation*}
Z_{3}\left( \ell \right) \sim \ell ^{\eta ^{\ast }-2}\,,
\end{equation*}%
so that we have: 
\begin{equation*}
Z_{3}\left( \ell \right) \underset{\ell \rightarrow 0}{\rightarrow }\infty
\,,
\end{equation*}%
provided that $\eta ^{\ast }<2$, which is the usual condition for a fixed
point to be a \textquotedblleft critical fixed point\textquotedblright\ \cite%
{6699}.

Consequently, the limit $\Lambda _{0}\rightarrow \infty $ of (\ref%
{eq:DeltaInf0}) implies the condition:%
\begin{equation}
\lim_{\Lambda _{0}\rightarrow \infty }\bar{\Delta}_{2}\left( \tilde{q}%
^{2},t\right) =\bar{\Delta}_{2}^{\infty }\left( \tilde{q}^{2}\right) =-%
\tilde{P}\left( \tilde{q}^{2}\right) \,,  \label{eq:DeltaInf}
\end{equation}%
so that both (\ref{eq:EMROWett1}) and (\ref{eq:RGflowGammaInt0}) greatly
simplify. However, after the limit of infinite cutoff $\Lambda _{0}$ has
been performed, it seems that the RG flow equation could display
singularities due to the term $1-\tilde{P}\left( \tilde{q}^{2}\right) \Gamma
_{\mathrm{int}}^{\left( 2\right) }\left[ \tilde{M}\right] $ appearing in the
denominator of a fraction. Taking benefit from the fact that the cutoff
function is actually an integral part of the theory, this difficulty can be
avoided if, as Ellwanger \cite{4436}, we consider the RG flow equation for
the \textsl{full} scale-de\-pen\-dent effective action $\Gamma \left[ \tilde{M},t%
\right] $ prior to taking the limit (\ref{eq:DeltaInf}).

\subsection{Back to full actions}

According to (\ref{eq:FullGamma0}) in which the rescaling and the
field-renormalization have been implemented (but not the limit $\Lambda
_{0}\rightarrow \infty $ yet), the full effective action defines as:%
\begin{equation}
\Gamma \left[ \tilde{M},t\right] =\frac{1}{2}\tilde{M}\cdot \bar{\Delta}%
_{2}^{-1}\cdot \tilde{M}+\Gamma _{\mathrm{int}}\left[ \tilde{M},t\right] \,.
\end{equation}%
Then, we may consider safely the limit (\ref{eq:DeltaInf}) within the
RG-flow equation for $\Gamma $ so that (\ref{eq:RGflowGammaInt0}) and (\ref%
{eq:EMROWett1}) leads respectively to (up to an additive constant):%
\begin{eqnarray}
\dot{\Gamma}\left[ \tilde{M},t\right] =\int_{q} &&q^{2}\frac{P^{\prime
}\left( q^{2}\right) }{\tilde{P}^{2}\left( \tilde{q}^{2}\right) }\left\{ 
\frac{1}{\,\Gamma ^{\left( 2\right) }\left[ \tilde{q}\right] }+\tilde{M}_{%
\tilde{q}}\,\,\tilde{M}_{-\tilde{q}}\right\}   \notag \\
&&+\mathcal{G}_{\text{\textrm{dil}}}\left( \Gamma ,\tilde{M},d_{\phi
}^{\left( cw\right) }\right) +\varpi _{0}\left( t\right) \mathcal{O}%
_{P}^{\infty }\left( \Gamma ,\tilde{M}\right) \,,  \label{eq:RGGammaPoint} \\
\mathcal{O}_{P}^{\infty }\left( \Gamma ,\tilde{M}\right) =-\int_{\tilde{q}}
&&\left\{ \tilde{P}^{-1}\left( \tilde{q}^{2}\right) \left[ \frac{1}{\Gamma
^{\left( 2\right) }\left[ \tilde{q};\tilde{M}\right] }+\tilde{M}_{\tilde{q}}%
\tilde{M}_{-\tilde{q}}\right] \right.   \notag \\
&&\left. +\tilde{M}_{\tilde{q}}\frac{\delta \Gamma }{\delta \tilde{M}_{%
\tilde{q}}}+1\right\} \,.  \label{eq:OPGam}
\end{eqnarray}

Because $\tilde{P}\left( \tilde{q}^{2}\right) $ is assumed to decrease
sufficiently rapidly toward zero when $\tilde{q}\rightarrow \infty $, a
supplementary redefinition of the field $\tilde{M}$ as (that, again, implies 
$n_{0}=0$):%
\begin{equation*}
\tilde{M}_{\tilde{q}}=\tilde{P}\left( \tilde{q}^{2}\right) \tilde{\Phi}_{%
\tilde{q}}\,,
\end{equation*}%
eliminates possible bad behavior of the equation for large $q$, and we
finally get:%
\begin{eqnarray}
\dot{\Gamma}\left[ \tilde{\Phi},t\right]  &=&\int_{q}q^{2}P^{\prime }\left(
q^{2}\right) \left( \frac{1}{\,\Gamma ^{\left( 2\right) }\left[ \tilde{q};%
\tilde{\Phi}\right] }+\tilde{\Phi}_{\tilde{q}}\,\,\tilde{\Phi}_{-\tilde{q}%
}\right)   \notag \\
&&+\mathcal{G}_{\text{\textrm{dil}}}\left( \Gamma ,\tilde{\Phi},d_{\phi
}^{\left( cw\right) }\right) +\varpi _{0}\left( t\right) \mathcal{\bar{O}}%
_{P}^{\infty }\left( \Gamma ,\tilde{\Phi}\right)   \notag \\
&&+\int_{q}\left[ \mathbf{q}\cdot \frac{\partial \ln P\left( \tilde{q}%
^{2}\right) }{\partial \mathbf{q}}\right] \tilde{\Phi}_{q}\frac{\delta
\Gamma }{\delta \tilde{\Phi}_{q}}\,, \\
\mathcal{\bar{O}}_{P}^{\infty }\left( \Gamma ,\tilde{\Phi}\right)  &=&-\int_{%
\tilde{q}}\left\{ \tilde{P}\left( \tilde{q}^{2}\right) \left[ \frac{1}{%
\Gamma ^{\left( 2\right) }\left[ \tilde{q};\tilde{\Phi}\right] }+\tilde{\Phi}%
_{\tilde{q}}\tilde{\Phi}_{-\tilde{q}}\right] +\tilde{\Phi}_{\tilde{q}}\frac{%
\delta \Gamma }{\delta \tilde{\Phi}_{\tilde{q}}}+1\right\} \,,
\end{eqnarray}%
which, once $\mathcal{\bar{O}}_{P}^{\infty }$ is split in two parts, also
reads:%
\begin{eqnarray}
\dot{\Gamma}\left[ \tilde{\Phi},t\right] &=&\int_{q}\left[ -\varpi
_{0}\left( t\right) P\left( q^{2}\right) +q^{2}P^{\prime }\left(
q^{2}\right) \right] \left\{ \frac{1}{\Gamma ^{\left( 2\right) }\left[ 
\tilde{q};\tilde{\Phi}\right] }+\tilde{\Phi}_{\tilde{q}}\tilde{\Phi}_{-%
\tilde{q}}\right\}  \notag \\
&&+\mathcal{G}_{\text{\textrm{dil}}}\left( \Gamma ,\tilde{M},d_{\phi
}^{\left( +\right) }\right) +\int_{q}\left[ \mathbf{q}\cdot \frac{\partial }{%
\partial \mathbf{q}}\ln \left\vert P\right\vert \right] \tilde{\Phi}_{q}%
\frac{\delta \Gamma }{\delta \tilde{\Phi}_{q}}\,.
\label{eq:RGGammaPointFinal}
\end{eqnarray}

Meanwhile the relation to $S_{\mathrm{int}}\left[ \tilde{\phi},t\right] $
has become:%
\begin{eqnarray}
S_{\mathrm{int}}\left[ \tilde{\phi},t\right] &=&-\tilde{\phi}\cdot \tilde{%
\Phi}-\frac{1}{2}\tilde{\phi}\cdot P^{-1}\cdot \tilde{\phi}+\Gamma \left[ 
\tilde{\Phi},t\right] \,, \\
P\cdot \tilde{\Phi} &=&-\tilde{\phi}+P\cdot \frac{\delta }{\delta \tilde{\phi%
}}S_{\mathrm{int}}\left[ \tilde{\phi},t\right] \,,
\end{eqnarray}%
so that, due to the relation (\ref{eq:DeltaInf}), the full action $S\left[ 
\tilde{\phi},t\right] $ may now be reconstructed\footnote{%
Contrary to the common procedure where the infinite limit of the overall
cutoff corresponded to (\ref{eq:DeltaInf0}) in which $Z_{3}\equiv 1$.} to
get the simple Legendre transformation (\ref{eq:001a}, \ref{eq:001b}) stated
in the introduction that with the properties:%
\begin{eqnarray*}
\dot{\Gamma}\left[ \tilde{\Phi},t\right] &=&\dot{S}\left[ \tilde{\phi},t%
\right] \,, \\
\mathcal{G}_{\text{\textrm{dil}}}\left( S,\tilde{\phi},d_{\phi }^{\left(
cw\right) }\right) &=&\mathcal{G}_{\text{\textrm{dil}}}\left( \Gamma ,\tilde{%
\Phi},d_{\phi }^{\left( cw\right) }\right) \,,
\end{eqnarray*}%
enables us to deduce (\ref{eq:RGGammaPointFinal}) directly from the
Wilsonian ERGE (\ref{eq:WilExtended}).

\section{Summary and final comments\label{Conc}}

After having presented a review of the various forms of ERGE currently
encountered in the litterature, with a particular attention given to the
ways in which RG-step 3 (field-renormalization) has been realized, we have
shown how an ERGE may be constructed in such a way that a simple Legendre
transformation relates the RG-flow equation for the full scale dependent
action $S\left[ \tilde{\phi},t\right] $ to that of the full scale dependent
effective action $\Gamma \left[ \tilde{\Phi},t\right] $. Though not new, the
first important ingredient is the reference to Wilson's achievement of
RG-step 1 via an \textquotedblleft incomplete integration\textquotedblright\
in which no reference to any explicit UV-cutoff is required. Indeed, the
number of degrees of freedom are reduced by integrating \textquotedblleft
more\textquotedblright\ the fields with large momenta that those with small
momenta. Clearly, in this view there is no overall UV-scale $\Lambda _{0}$
but only a pure momentum-scale of reference, at least in the RG-flow
equation of the full action $S$ (Wilsonian ERGE). Qualitatively, the
situation may be compared to the perturbation field theory dimensionally regularized that avoids any
consideration of explicit cutoff procedure but where a pure momentum scale
automatically appears for dimensional reasons. Again not new, the
second important ingredient is the systematic inclusion of the cutoff
quadratic form into the full scale-de\-pen\-dent effective action $\Gamma $,
that eliminates all the reference to any explicit cutoff procedure within
the RG-flow equation. Then, provided the usual limit of infinite overall
UV-cutoff $\Lambda _{0}\rightarrow \infty $ is undertaken, it only remains a
formal running momentum scale (and eventual field-redefinitions to make the
ERGEs well defined). Consequently the \textsl{full} actions no longer need
to be affected by the introduction of arbitrary cutoff functions of any kind
contrary to the common practice. It remains to link the two flow equations
for the full actions, what was commonly thought to be impossible. To this
regard, the most important (the third) ingredient is the global
implementation of RG-step 3 in the ERGE, via the \textquotedblleft
operator\textquotedblright\ responsible for an infinitesimal change of
normalization of the field by a pure constant (EMRO). Indeed with an EMRO,
RG-step 3 may be seen as being obligatory \textquotedblleft \textsl{equally
implemented}\textquotedblright\ on all the fields $\phi _{q}$ with $0\leq
\left\vert q\right\vert <\infty $. This leaves no room for any explicit
cutoff. If, nevertheless, one insists to consider explicit cutoffs (IR and
UV) and try to maintain the recourse to an EMRO, then RG-step 3 appears to
be equally implemented only in the finite range $\mu \leq \left\vert
q\right\vert <\Lambda _{0}$ with $\mu $ an effective overall IR cutoff
linked to the usual overall UV-cutoff $\Lambda _{0}$ so that $\mu /\Lambda
=\Lambda /\Lambda _{0}$ ($\Lambda $ being the running momentum-scale). Then,
the usual limit $\Lambda _{0}\rightarrow \infty $ of the common view is
linked to the limit $\mu \rightarrow 0$. Using the common procedure that
refers to the vicinity of a fixed point to justify the limit $\Lambda
_{0}\rightarrow \infty $, we show that, provided the anomalous dimension
satisfies the condition $\eta ^{\ast }<2$, the limit $\mu \rightarrow 0$
simplifies the usual relation between the UV and IR-cutoff functions so that
a simple Legendre transformation links the two RG-flow equations.

The proof of existence of such a Legendre transformation has been given
following the traditional way that first considers explicit cutoff functions
so that the scale dependent actions $S\left[ \tilde{\phi},t\right] $ and $%
\Gamma \left[ \tilde{\Phi},t\right] $ have the usual definitions. In
particular $\Gamma \left[ \tilde{\Phi},t\right] $ may continue to be seen as
interpolating between a \textquotedblleft bare\textquotedblright\ action for
small $t$ and the \textquotedblleft usual\textquotedblright\ effective
action when $t\rightarrow \infty $. However, the main interest for field
theory remains the vicinity of fixed points $\Gamma ^{\ast }\left[ \tilde{%
\Phi}\right] $ which are a priori unknown prior to any explicit calculations.

We hope that the present work be of some help in the treatment of modern
problems of field theory in which local symmetries (incompatible with an
explicit cutoff procedure) are involved.

\appendix{}

\section{Field-renormalization and anomalous dimension\label{defZ}}

In \cite{5744}, a confusion has been made between the field-renormalization
function $Z_{3}$ and its inverse. This is why we present again (see also 
\cite{2727}) the arguments that relate $Z_{3}\left( \ell \right) $ to the
anomalous dimension $\eta ^{\ast }$ --and by extension to $\eta \left(
t\right) $-- as indicated by (\ref{eq:Z3}, \ref{eq:cdet}).

Provided one considers sufficiently large distances (small momenta), the
original two-point correlation function: 
\begin{equation*}
G\left( \left\vert q_{1}\right\vert ,S\right) \delta \left(
q_{1}+q_{2}\right) =\left\langle \phi _{q_{1}}\phi _{q_{2}}\right\rangle
_{S}\,,
\end{equation*}%
is preserved after renormalization. The decimation RG-step 1 does not modify
the physics at large distances ($q_{1}\rightarrow 0$), then only the two
remaining steps ( rescaling and field-renormalization) can modify the
property of the two point correlation function. Field-renormalization yields:%
\begin{equation*}
\left\langle \phi _{q_{1}}\phi _{q_{2}}\right\rangle _{S}=Z_{3}\left\langle
\phi _{q_{1}}^{R}\phi _{q_{2}}^{R}\right\rangle _{\bar{S}}\,,
\end{equation*}%
and the supplementary rescaling implies\footnote{%
Roughly speaking we have $\left\vert q\right\vert <\mu $ and after rescaling 
$\left\vert q^{\prime }\right\vert <\Lambda $ with $\mu =\ell \Lambda $,
thus $\left\vert q\right\vert =\ell \left\vert q\,^{\prime }\right\vert $ ,
and by definition of $\bar{d}_{\phi }^{\left( c\right) }$, $\tilde{\phi}%
_{\ell \tilde{q}_{1}^{\prime }}=\ell ^{2\bar{d}_{\phi }^{\left( c\right) }}%
\tilde{\phi}_{\tilde{q}_{1}^{\prime }}$.\label{lq}}:%
\begin{equation}
\left\langle \tilde{\phi}_{\tilde{q}_{1}}\tilde{\phi}_{\tilde{q}%
_{2}}\right\rangle _{S}=Z_{3}\ell ^{2\bar{d}_{\phi }^{\left( c\right)
}}\left\langle \tilde{\phi}_{\tilde{q}_{1}^{\prime }}^{R}\tilde{\phi}_{%
\tilde{q}_{2}^{\prime }}^{R}\right\rangle _{\bar{S}}\,,  \label{eq:rel}
\end{equation}%
in which $\bar{d}_{\phi }^{\left( c\right) }$ means $d_{\phi }^{\left(
c\right) }-d$ and $d_{\phi }^{\left( c\right) }$ is the classical dimension
of $\phi \left( x\right) $ [defined by (\ref{eq:dc})].

(In the following of this appendix we assume that all quantities are
dimensionless and we forget the tilde.)

Expressed on $G\left( q,S\right) $ --after having taken into account the
dimension of the $\delta $-function-- (\ref{eq:rel}) becomes:%
\begin{equation}
G\left( \left\vert q\right\vert ,S\right) =Z_{3}\ell ^{2d_{\phi }^{\left(
c\right) }-d}G\left( \left\vert q^{\prime }\right\vert ,S\right) \,.
\label{eq:rel1}
\end{equation}

In general (\ref{eq:rel1}) is a complicated relation. But at a fixed point $%
S^{\ast }=\bar{S}^{\ast }$ one expects a specific momentum dependence of the
correlation function $G^{\ast }\left( q\right) \equiv G\left( q,S^{\ast
}\right) $.

Indeed, knowing that $\left\vert q\right\vert =\ell $\thinspace $\left\vert
q^{\prime }\right\vert $ (see footnote \ref{lq}), we get at a fixed point: 
\begin{equation}
G^{\ast }\left( \left\vert q\right\vert \right) =Z_{3}\ell ^{2d_{\phi
}^{\left( c\right) }-d}G^{\ast }\left( \frac{\left\vert q\right\vert }{\ell }%
\right) \,,  \label{dzeta2}
\end{equation}%
which implies that $G^{\ast }\left( \left\vert q\right\vert \right) \propto
\left\vert q\right\vert ^{\alpha }$. This is precisely what is physically
expected at a critical point with:%
\begin{equation}
G\left( \left\vert q\right\vert \right) \underset{q\rightarrow 0}{\sim }%
G_{0}\left\vert q\right\vert ^{\eta ^{\ast }-2}\,,  \label{Gasy}
\end{equation}%
in which $\eta ^{\ast }$ is called the anomaous dimension of the field.

When reported in eq. (\ref{dzeta2}) this gives:%
\begin{equation*}
Z_{3}\ell ^{2d_{\phi }^{\left( c\right) }-d}\ell ^{2-\eta ^{\ast }}=1\,,
\end{equation*}%
and thus, taking into account (\ref{eq:dc}):%
\begin{equation}
Z_{3}=\ell ^{2\left( \eta ^{\ast }-1+n_{0}\right) }\,,  \label{eq:PowerLaw}
\end{equation}%
which is the fixed point expression of (\ref{eq:Z3}, \ref{eq:cdet}) which,
in turn, is nothing but an extension, away from any fixed point, of the
power law (\ref{eq:PowerLaw}).

\section{Obtaining the expression of a simple EMRO\label{ObtainEMRO}}

In this appendix all quantities are dimensionless.

With a view to use it in the structural method described in section \ref%
{Prince}, we aim at obtaining a simple expression of an EMRO associated to a
given Wilsonian ERGE. To this end, we follow the procedure developped by
O'Dwyer and Osborn in their appendix D of \cite{6228} with the difference
that we deal with the full action $S$ and a non-linear implementation of
RG-step 3 instead of a \textquotedblleft modified\textquotedblright\
Polchinski ERGE (for $S_{\mathrm{int}}$) and a RG-step 3 effected linearly
(see section \ref{modif}).

We start by considering a Wilsonian ERGE of the following general form:%
\begin{equation}
\dot{S}=\int_{q}\left[ G\left( q^{2}\right) \left( \frac{\delta ^{2}S}{%
\delta \phi _{q}\delta \phi _{-q}}-\frac{\delta S}{\delta \phi _{q}}\frac{%
\delta S}{\delta \phi _{-q}}\right) +H\left( q^{2}\right) \phi _{q}\frac{%
\delta S}{\delta \phi _{q}}\right] +\mathcal{G}_{\mathrm{dil}}\left(
S,d_{\phi }\right) \,,  \label{B1}
\end{equation}%
in which nothing is said on whether RG-step 3 has or has not been
implemented. The question is to determine an EMRO $\mathcal{O}_{0}\left(
S^{\ast },G,H,d_{\phi }\right) $ --i.e., a redundant operator with a zero
eigenvalue-- associated to that ERGE.

Let us consider the linearization of the ERGE (\ref{B1}) in the vicinity of
a fixed point $S^{\ast }$, it leads to the eigenvalue equation:%
\begin{equation}
\mathcal{D\,U}^{\ast }=\lambda \,\mathcal{U}^{\ast }\,,  \label{eq:VP}
\end{equation}%
with:%
\begin{eqnarray}
\mathcal{D} &=&\mathcal{D}_{1}+\mathcal{D}_{2}+\mathcal{D}_{3}+\mathcal{D}%
_{4}\,,  \label{eq:Dbis} \\
\mathcal{D}_{1} &=&\int_{q}\left[ \left( d-d_{\phi }\right) \,\phi _{q}+%
\mathbf{q}\cdot \frac{\partial }{\partial \mathbf{q}}\phi _{q}\right] \,%
\frac{\delta }{\delta \phi _{q}}\,,  \label{eq:D1} \\
\mathcal{D}_{2} &=&\int_{q}G\left( q^{2}\right) \frac{\delta ^{2}}{\delta
\phi _{q}\delta \phi _{-q}}\,,  \label{eq:D2} \\
\mathcal{D}_{3} &=&\int_{q}H\left( q^{2}\right) \phi _{q}\frac{\delta }{%
\delta \phi _{q}}\,,  \label{eq:D3} \\
\mathcal{D}_{4} &=&-2\int_{q}G\left( q^{2}\right) \frac{\delta S^{\ast }}{%
\delta \phi _{q}}\frac{\delta }{\delta \phi _{-q}}\,.  \label{eq:D4}
\end{eqnarray}

From the general form of a redundant operator:%
\begin{equation}
\mathcal{O}_{\Psi }^{\ast }=\int_{q}\left[ \Psi _{q}\frac{\delta S^{\ast }}{%
\delta \phi _{q}}-\frac{\delta \Psi _{q}}{\delta \phi _{q}}\right] \,,
\label{eq:OPsiBis}
\end{equation}%
and the definitions (\ref{eq:D1}--\ref{eq:D4}), following similar
calculations as those presented in appendix D of \cite{6228}, one can show
that: 
\begin{equation}
\mathcal{D\,}O_{\Psi }^{\ast }=O_{\mathcal{D}_{1}\Psi }^{\ast }\,,
\label{eq:DOPsiBis}
\end{equation}%
in which:%
\begin{equation}
\mathcal{D}_{1}\,\Psi _{q}=\mathcal{D\,}\Psi _{q}+\left[ \left( d_{\phi
}-d-H\left( q^{2}\right) \right) -\mathbf{q}\cdot \frac{\partial }{\partial 
\mathbf{q}}\right] \Psi _{q}\,.  \label{eq:DBarPsiBis}
\end{equation}

As noted in \cite{6228} this result demonstrates that the operators $%
\mathcal{O}_{\Psi }^{\ast }$ form a closed subspace under RG flow near a
fixed point \cite{4011,2835}.

In order to construct an EMRO, the idea of O'Dwyer and Osborn is to look for
an operator $\mathcal{O}_{\bar{\Psi}}^{\ast }$ with:%
\begin{equation}
\bar{\Psi}_{q}=Q\left( q^{2}\right) \,\mathcal{F}\left( q^{2}\right) \,,
\label{eq:PsiQFBis}
\end{equation}%
in which $\mathcal{F}\left( q^{2}\right) $ is assumed to have the property:%
\begin{equation}
\mathcal{D\,F}\left( q\right) =\left( \mathbf{q}\cdot \frac{\partial }{%
\partial \mathbf{q}}+\lambda _{\mathcal{F}}\right) \,\mathcal{F}\left(
q\right) \,.  \label{eq:eqF}
\end{equation}

Consequently (\ref{eq:DBarPsiBis}) gives:%
\begin{equation*}
\mathcal{D}_{1}\bar{\Psi}_{q}=\left( d_{\phi }-d+\lambda _{\mathcal{F}%
}\right) \bar{\Psi}_{q}-\left( 2q^{2}\frac{Q^{\prime }}{Q}+H\left(
q^{2}\right) \right) \bar{\Psi}_{q}\,,
\end{equation*}%
so that choosing $Q\left( x\right) $ as defined by:%
\begin{equation}
2x\frac{Q^{\prime }\left( x\right) }{Q\left( x\right) }=-H\left( x\right) \,,
\label{eq:RelQH}
\end{equation}%
implies that:%
\begin{equation*}
\mathcal{D}_{1}\,\mathcal{O}_{\bar{\Psi}}^{\ast }=\left( d_{\phi
}-d\,+\lambda _{\mathcal{F}}\right) \mathcal{O}_{\bar{\Psi}}^{\ast }\,,
\end{equation*}%
and, owing to (\ref{eq:DOPsiBis}):%
\begin{equation*}
\mathcal{D\,O}_{\bar{\Psi}}^{\ast }=\left( d_{\phi }-d\,+\lambda _{\mathcal{F%
}}\right) \mathcal{O}_{\bar{\Psi}}^{\ast }\,,
\end{equation*}%
so that for:%
\begin{equation}
\lambda _{\mathcal{F}}=d-d_{\phi }\,,  \label{eq:LambdaF}
\end{equation}%
$\mathcal{O}_{\bar{\Psi}}^{\ast }$ is an EMRO.

To construct $\mathcal{F}\left( q^{2}\right) $ with the required properties,
O'Dwyer and Osborn propose the following form:%
\begin{equation*}
\mathcal{F}\left( q^{2}\right) =a\left( q^{2}\right) \,\phi _{q}+b\left(
q^{2}\right) \,\frac{\delta S^{\ast }}{\delta \phi _{-q}}\,.
\end{equation*}

So that, using the properties --which follow from the definition of $%
\mathcal{D}$ [eqs. (\ref{eq:Dbis}, \ref{eq:D4})]:%
\begin{eqnarray}
\mathcal{D\,}\phi _{q} &=&\left( d-d_{\phi }+H\left( q^{2}\right) +\mathbf{q}%
\cdot \frac{\partial }{\partial \mathbf{q}}\right) \phi _{q}-2G\left(
q^{2}\right) \frac{\delta S^{\ast }}{\delta \phi _{-q}}\,,
\label{eq:dphiqbis} \\
\mathcal{D\,}\frac{\delta S^{\ast }}{\delta \phi _{q}} &=&\left[ \left(
d_{\phi }-H\left( q^{2}\right) \right) \frac{\delta }{\delta \phi _{q}}%
\,+\left( \mathbf{q}\cdot \frac{\partial }{\partial \mathbf{q}}\frac{\delta 
}{\delta \phi _{q}}\right) \right] S^{\ast }\,,  \label{eq:dsdphibis}
\end{eqnarray}%
it is not too complicated to show that the conditions (\ref{eq:eqF}--\ref%
{eq:LambdaF}) lead to the following coupled differential equations for $%
a\left( x\right) $ and $b\left( x\right) $ (in which the prime $^{\prime }$
stands for $\frac{d}{dx}$):%
\begin{eqnarray}
H\left( x\right) a\left( x\right) -2xa^{\prime }\left( x\right) &=&0\,,
\label{eq:a} \\
2a\left( x\right) G\left( x\right) +b\left( x\right) \left( d-2d_{\phi
}\right) +H\left( x\right) b\left( x\right) +2xb^{\prime }\left( x\right)
&=&0\,.  \label{eq:b}
\end{eqnarray}

Actually, (\ref{eq:a}) is similar to (\ref{eq:RelQH}) and we may thus set:%
\begin{equation}
a\left( x\right) =\frac{A}{Q\left( x\right) }\,,  \label{eq:sola}
\end{equation}%
in which $A$ is some constant depending on initial conditions.

Then, if we introduce the function $B\left( x\right) $:%
\begin{equation*}
B\left( x\right) =-Q\left( x\right) b\left( x\right) \,,
\end{equation*}%
and choose $A=1$, using (\ref{eq:RelQH}) one may show that the differential
equation (\ref{eq:b}) becomes \cite{7849}:%
\begin{eqnarray}
G\left( x\right) -\varpi _{g}B\left( x\right) -H\left( x\right) B\left(
x\right) -xB^{\prime }\left( x\right)  &=&0\,,  \label{eq:Binit} \\
\varpi _{g} &=&\frac{d}{2}-d_{\phi }.  \label{eq:omegaBarg}
\end{eqnarray}

Assuming the initial condition:%
\begin{equation*}
B\left( 0\right) =1\,,
\end{equation*}%
the solution of (\ref{eq:Binit}) then reads:%
\begin{eqnarray}
B\left( x\right)  &=&\lim_{\epsilon \rightarrow 0}\left\{ \left( \frac{%
\epsilon }{x}\right) ^{\varpi _{g}}\frac{1}{C_{\epsilon }\left( x\right) }%
\left[ 1+\epsilon ^{-\varpi _{g}}\int_{\epsilon }^{x}u^{\varpi
_{g}-1}G(u)C_{\epsilon }\left( u\right) du\right] \right\} \,,  \label{Bsol}
\\
C_{\epsilon }\left( x\right)  &=&e^{\int_{\epsilon }^{x}\frac{H(u)}{u}%
\,du}\,.  \label{Csol}
\end{eqnarray}

Before discussing these expressions with explicit examples, let us come back
to the differential equation (\ref{eq:Binit}).

Let us assume that we know a solution $B_{0}\left( x\right) $ of (\ref%
{eq:Binit}) and that we use it to modify the ERGE (\ref{B1}) such that:%
\begin{eqnarray*}
\dot{S} &\rightarrow &\dot{S}+\alpha \,\mathcal{O}_{\bar{\Psi}_{0}} \,, \\
\bar{\Psi}_{0,q} &=&\phi _{q}-B_{0}\left( q^{2}\right) \frac{\delta S^{\ast }%
}{\delta \phi _{-q}}\,,
\end{eqnarray*}%
then (\ref{eq:Binit}) is changed into a differential equation for $B_{\alpha
}\left( x\right) $ which reads:

\begin{equation*}
0=G\left( x\right) +\alpha \left[ B_{0}\left( x\right) -B_{\alpha }\left(
x\right) \right] -\varpi _{g}B_{\alpha }\left( x\right) -H\left( x\right)
B_{\alpha }\left( x\right) -xB_{\alpha }^{\prime }\left( x\right) \,,
\end{equation*}%
so that $B_{\alpha }\left( x\right) =B_{0}\left( x\right) $ is still a
solution. Consequently, an EMRO like $\mathcal{O}_{\bar{\Psi}_{0}}$
(corresponding to $\varpi _{g}=0$) may be used to implement RG-step 3
without altering its quality of EMRO.

Let us illustrate this on two examples on which we shall show that the
initial condition $B_{\alpha }\left( 0\right) =\mathrm{const.}$, imposed by
the requirement of quasi-locality for $S$, induces the condition $\eta
^{\ast }<2$.

The two examples correspond to a choice of cutoff function (\ref{eq:defP})
with, respectively, $n_{0}=0$ (Wilson's choice) and $n_{0}=1$ (Polchinski's
choice).

\subsection{Wilson's choice}

The Wilson ERGE extended to an arbitrary cutoff function $P\left(
q^{2}\right) $ as given by (\ref{eq:defP}) with $n_{0}=0$ and prior to
realization of RG-step 3, corresponds to (\ref{B1}) with the following
choices [see eqs. (\ref{eq:WilExtended}, \ref{eq:dcwil}) with $\varpi _{0}=0$%
]:%
\begin{eqnarray*}
G\left( x\right)  &=&-xP^{\prime }\left( x\right) \,, \\
H\left( x\right)  &=&-2x\frac{P^{\prime }\left( x\right) }{P\left( x\right) }%
\,, \\
d_{\phi } &=&d_{\phi }^{\left( cw\right) }=\frac{d}{2}\,, \\
\varpi _{g} &=&0\,,
\end{eqnarray*}%
and the solution (\ref{Bsol}, \ref{Csol}) reads:%
\begin{eqnarray}
C_{\epsilon }\left( x\right)  &=&\left( \frac{P\left( \epsilon \right) }{%
P\left( x\right) }\right) ^{2}\,,  \label{Csol0} \\
B\left( x\right)  &=&\lim_{\epsilon \rightarrow 0}\left\{ P^{2}\left(
x\right) \left[ \frac{1}{P^{2}\left( \epsilon \right) }-\int_{\epsilon }^{x}%
\frac{P^{\prime }\left( u\right) }{P^{2}\left( u\right) }du\right] \right\}
\,,  \label{Bsol0}
\end{eqnarray}%
which, after integration and provided that $P\left( 0\right) =1$, give:%
\begin{equation*}
B\left( x\right) =P\left( x\right) \,.
\end{equation*}

The second step, consists in considering the case $\varpi =\varpi
_{0}=1-\eta ^{\ast }/2$ in (\ref{B1}). That amounts to finding the solution
of (\ref{eq:Binit}) in which the following changes are effected:%
\begin{eqnarray*}
G\left( x\right)  &\rightarrow &G\left( x\right) +\varpi _{0}B\left(
x\right) \,, \\
\varpi _{g} &\rightarrow &\varpi _{g}+\varpi _{0}\,,
\end{eqnarray*}%
in that case (\ref{Csol0}) is unchanged and (\ref{Bsol0}) is replaced by:%
\begin{eqnarray*}
B\left( x\right) =\lim_{\epsilon \rightarrow 0} &&\left\{ \left( \frac{%
\epsilon }{x}\right) ^{\varpi _{0}}P^{2}\left( x\right) \left[ \frac{1}{%
P^{2}\left( \epsilon \right) }+\varpi _{0}\epsilon ^{-\varpi
_{0}}\int_{\epsilon }^{x}u^{\varpi _{0}-1}\frac{B\left( u\right) }{%
P^{2}\left( u\right) }du\right. \right.  \\
&&\left. \,\left. -\epsilon ^{-\varpi _{0}}\int_{\epsilon }^{x}u^{\varpi
_{0}}\frac{P^{\prime }\left( u\right) }{P^{2}\left( u\right) }du\right]
\right\} \,.
\end{eqnarray*}

By integrating by parts the last term, it comes, after some rearrangement:%
\begin{eqnarray*}
B\left( x\right) =\lim_{\epsilon \rightarrow 0} &&\left\{ P\left( x\right)
+\left( \frac{\epsilon }{x}\right) ^{\varpi _{0}}P^{2}\left( x\right) \left( 
\frac{1}{P^{2}\left( \epsilon \right) }-\frac{1}{P\left( \epsilon \right) }%
\right) \right.  \\
&&\left. +\left( \frac{1}{x}\right) ^{\varpi _{0}}P^{2}\left( x\right)
\varpi _{0}\int_{\epsilon }^{x}u^{\varpi _{0}-1}\left( \frac{B\left(
u\right) }{P^{2}\left( u\right) }-\frac{1}{P\left( u\right) }\right)
du\right\} \,,
\end{eqnarray*}%
in which the second term gives $0$ in the limit $\epsilon \rightarrow 0$
provided that $\varpi _{0}>0$ (and $P\left( 0\right) =\mathrm{const}$) and
the second term vanishes if $B\left( x\right) =P\left( x\right) $ which is
actually the case, what confirms (\ref{eq:PsiPS}).

We note that the condition $\varpi _{0}>0$ corresponds to the usual
condition $\eta ^{\ast }<2$ to have a \textquotedblleft critical fixed
point\textquotedblright\ (see \cite{6699}).

\subsection{Polchinski's choice}

The Polchinski choice for the arbitrary cutoff function $P\left(
q^{2}\right) $ is (\ref{eq:defP}) with $n_{0}=1$. This choice, which is
perfectly acceptable, corresponds to (\ref{B1}) with:%
\begin{eqnarray*}
G\left( x\right)  &=&-K^{\prime }\left( x\right) \,, \\
H\left( x\right)  &=&-2x\frac{K^{\prime }\left( x\right) }{K\left( x\right) }%
\,, \\
d_{\phi } &=&d_{\phi }^{\left( c\right) }=\frac{d}{2}-1\,, \\
\varpi _{g} &=&1\,,
\end{eqnarray*}%
and the solution (\ref{Bsol}, \ref{Csol}) reads:%
\begin{eqnarray}
C_{\epsilon }\left( x\right)  &=&\left( \frac{K\left( \epsilon \right) }{%
K\left( x\right) }\right) ^{2}\,,  \label{Csol1} \\
B\left( x\right)  &=&\lim_{\epsilon \rightarrow 0}\left\{ \frac{K^{2}\left(
x\right) }{x}\left[ \frac{\epsilon }{K^{2}\left( \epsilon \right) }%
-\int_{\epsilon }^{x}\frac{K^{\prime }(u)}{K^{2}\left( u\right) }du\right]
\right\} \,,  \label{Bsol1}
\end{eqnarray}%
then, after integration, we get in the limit $\epsilon \rightarrow 0$ (and
provided that $K\left( 0\right) =1$)%
\begin{equation}
B\left( x\right) =\frac{K\left( x\right) }{x}\left[ 1-K\left( x\right) %
\right] \,.  \label{BPol}
\end{equation}

The second step, consists in considering the case $\varpi =\varpi _{1}=-\eta
^{\ast }/2$ in (\ref{B1}) that corresponds to setting $\varpi _{g}=1-\eta
^{\ast }/2=\varpi _{0}$ and $G\left( x\right) \rightarrow G\left( x\right)
+\varpi _{1}B\left( x\right) $. As previously, (\ref{Csol1}) is unchanged
and (\ref{Bsol1}) is replaced by:%
\begin{eqnarray*}
B\left( x\right) =\lim_{\epsilon \rightarrow 0} &&\left\{ \left( \frac{%
\epsilon }{x}\right) ^{\varpi _{0}}K^{2}\left( x\right) \left[ \frac{1}{%
K^{2}\left( \epsilon \right) }+\varpi _{1}\epsilon ^{-\varpi
_{0}}\int_{\epsilon }^{x}u^{\varpi _{0}-1}\frac{B\left( u\right) }{%
K^{2}\left( u\right) }du\right. \right.  \\
&&\left. \left. -\epsilon ^{-\varpi _{0}}\int_{\epsilon }^{x}u^{\varpi
_{0}-1}\frac{K^{\prime }\left( u\right) }{K^{2}\left( u\right) }du\right]
\right\} \,,
\end{eqnarray*}%
By integrating by parts the last term, it comes, after some rearrangement
(using the relation $\varpi _{1}=\varpi _{0}-1$):%
\begin{eqnarray*}
B\left( x\right) =\lim_{\epsilon \rightarrow 0} &&\left\{ \frac{K\left(
x\right) }{x}+\left( \frac{\epsilon }{x}\right) ^{\varpi _{0}}K^{2}\left(
x\right) \left( \frac{1}{K^{2}\left( \epsilon \right) }-\frac{1}{\epsilon
K\left( \epsilon \right) }\right) \right.  \\
&&\left. +\varpi _{1}\frac{K^{2}\left( x\right) }{x^{\varpi _{0}}}%
\int_{\epsilon }^{x}u^{\varpi _{0}-1}\left( \frac{B\left( u\right) }{%
K^{2}\left( u\right) }-\frac{1}{u\,K\left( u\right) }\right) du\right\} \,,
\end{eqnarray*}%
if one uses (\ref{BPol}) in the r.h.s. then the last term simplifies and we
get:%
\begin{eqnarray*}
B\left( x\right) =\lim_{\epsilon \rightarrow 0} &&\left\{ \frac{K\left(
x\right) }{x}+\left( \frac{\epsilon }{x}\right) ^{\varpi _{0}}K^{2}\left(
x\right) \left( \frac{1}{K^{2}\left( \epsilon \right) }-\frac{1}{\epsilon
K\left( \epsilon \right) }\right) \right.  \\
&&\left. -\varpi _{1}\frac{K^{2}\left( x\right) }{x^{\varpi _{0}}}%
\int_{\epsilon }^{x}u^{\varpi _{0}-2}du\right\} \,,
\end{eqnarray*}%
which, after integration gives:%
\begin{equation*}
B\left( x\right) =\lim_{\epsilon \rightarrow 0}\left\{ \left[ \frac{K\left(
x\right) }{x}\left( 1-K\left( x\right) \right) +\left( \frac{\epsilon }{x}%
\right) ^{\varpi _{0}}K^{2}\left( x\right) \left( \frac{1}{K^{2}\left(
\epsilon \right) }-\frac{1}{\epsilon K\left( \epsilon \right) }+\frac{1}{%
\epsilon }\right) \right] \right\} \,,
\end{equation*}%
and the last term vanishes in the limit $\epsilon \rightarrow 0$ provided
that $K\left( 0\right) =1$ and $\varpi _{0}>0$ (i.e. $\eta ^{\ast }>2$). We
thus find (\ref{BPol}).

We note that the EMRO is not as simple as in the preceding case but, once
extended out of the fixed point, it may well be used to implement RG-step 3
within an ERGE to give, instead of (\ref{eq:WilModif}) for $n_{0}=1$: 
\begin{eqnarray}
\dot{S}\left[ \phi ,t\right]  &=&-\int_{q}K^{\,\prime }\left( q^{2}\right) %
\left[ \frac{\delta ^{2}S}{\delta \phi _{q}\delta \phi _{-q}}-\frac{\delta S%
}{\delta \phi _{q}}\frac{\delta S}{\delta \phi _{-q}}+2q^{2}K^{-1}\phi _{q}%
\frac{\delta S}{\delta \phi _{q}}\right]   \notag \\
&&+\mathcal{G}_{\mathrm{dil}}\left( S,\phi ,d_{\phi }^{\left( c\right)
}\right) \,+\varpi _{1}\left( t\right) \mathcal{O}_{K}\,,
\label{eq:WilStruct2}
\end{eqnarray}%
in which:%
\begin{eqnarray*}
d_{\phi }^{\left( c\right) } &=&\frac{d-2}{2}\,,\quad \varpi _{1}\left(
t\right) =-\frac{\eta \left( t\right) }{2}\,,\quad K\left( 0\right) =1\,, \\
\mathcal{O}_{K} &=&\int_{q}\left[ \frac{K\left( q^{2}\right) }{q^{2}}\left[
1-K\left( q^{2}\right) \right] \left( \frac{\delta ^{2}S}{\delta \phi
_{q}\delta \phi _{-q}}-\frac{\delta S}{\delta \phi _{q}}\frac{\delta S}{%
\delta \phi _{-q}}\right) +\phi _{q}\frac{\delta S}{\delta \phi _{q}}\right]
\,.
\end{eqnarray*}

$\mathcal{O}_{K}^{\ast }$ being an EMRO for (\ref{eq:WilStruct2}).

\end{document}